\documentclass[a4paper,12pt]{article}

\usepackage{amsmath}
\usepackage{amssymb}
\usepackage{fullpage}
\usepackage{graphicx}
\usepackage{natbib}
\usepackage{amsthm}

\newcommand{\braces}[1]{\set{#1}}
\newcommand{\brackets}[1]{\left[#1\right]}

\newcommand{\E}[1]{\mathbb{E}\left[#1\right]}
\newcommand{\iid}{\ensuremath{\mathrm{i.i.d.}}}

\newcommand{\rmd}{\ensuremath{\,\mathrm{d}}}
\newcommand{\set}[1]{\left\{#1\right\}}

\renewcommand{\hat}{\widehat}
\newcommand{\tr}{^{\top}}

\newtheorem{prop}{Proposition}
\newtheorem{rem}{Remark}

\usepackage{hyperref}
\begin{document}
\title{Estimating the Hawkes process from a discretely observed sample
  path}
\author{Feng Chen, Jeffrey Kwan, and Tom Stindl\\School of Mathematics and Statistics \\UNSW Sydney}
\maketitle
\begin{abstract}
  The Hawkes process is a widely used model in many areas, such as
  finance, seismology, neuroscience, epidemiology, and social
  sciences. Estimation of the Hawkes process from continuous
  observations of a sample path is relatively straightforward using
  either the maximum likelihood or other methods. However, estimating
  the parameters of a Hawkes process from observations of a sample
  path at discrete time points only is challenging due to the
  intractability of the likelihood with such data. In this work, we
  introduce a method to estimate the Hawkes process from a discretely
  observed sample path. The method takes advantage of a state-space
  representation of the incomplete data problem and use the sequential
  Monte Carlo (aka particle filtering) to approximate the likelihood
  function. As an estimator of the likelihood function the SMC
  approximation is unbiased, and therefore it can be used together
  with the Metropolis-Hastings algorithm to construct Markov Chains to
  approximate the likelihood distribution, or more generally, the
  posterior distribution of model parameters. The performance of the
  methodology is assessed using simulation experiments and compared
  with other recently published methods. The proposed estimator is
  found to have a smaller mean square error than the two benchmark
  estimators. The proposed method has the additional advantage that
  confidence intervals for the parameters are easily available. We
  apply the proposed estimator to the analysis of weekly count data on
  measles cases in Tokyo Japan and compare the results to those by
  one of the benchmark methods.
\end{abstract}

\section{Introduction}
\label{sec:intro}

In many applications that involve sequences of events such as
earthquakes, infectious disease transmission in a community, and
financial transactions, a frequently noted phenomenon is the temporal
clustering of the events. A widely used statistical model for event
sequences with this clustering feature is the Hawkes process
\citep{Hawkes1971}. Fitting the Hawkes process has most often been
done by the maximum likelihood (ML) method \citep{Ogata1978,Ozaki1979}, which
involves evaluating the logarithm of the likelihood of the model
relative to a continuously observed sample path up to a censoring time
and then maximising it as a function of the model parameters to be
estimated. The Expectation-Maximisation (EM) algorithm has also been
used to as a numerically more stable alternative to obtain the ML
estimator of the Hakes process, especially in the multivariate case
\citep{Chornoboy1988}.

When the sample path of the Hawkes process is not continuously
monitored, or when the event times are recorded with limited
precision, then the data available consists only of the values of the
sample path at isolated observation time points or equivalently,
counts of events in disjoint time intervals. However, the likelihood
of the Hawkes process with such data is difficult to compute in
general, so fitting the Hawkes process with such data is
challenging. Recently, there have been attempts to address this
challenge in the literature. \cite{Cheysson2022} proposed a spectral
method, namely the maximum Whittle likelihood method, to estimate the
Hawkes process with count data. They also established the consistency
and asymptotic normality of the estimator. However, they did not
provide an estimator for the standard errors of the
estimator. Moreover, their method relies on the stationarity of the
count time series and only works on data with regularly spaced
observation time points and Hawkes process models with a constant
background intensity.

\cite{Shlomovich2022jcgs} proposed a Monte Carlo
Expectation-Maximization (MCEM) algorithm to estimate the parameters of
the Hawkes process with count data, whereby the conditional
expectation of the complete data loglikelihood in the Expectation-
(E-) step of an EM cycle is approximated by a Monte Carlo (MC)
method. They also applied the method to estimation of the multivariate
Hawkes process from multi-type event count data
\citep{Shlomovich2022}. While their method does not require regularly
spaced observation time points in the data or a constant background
intensity in the Hawkes process model, due to their specific choice of
the Monte Carlo sample generation method, the resulting MC estimate of
the conditional expectation of the complete-data log-likelihood in the
E-step is biased in general, and the bias seems to be inherited by the
estimator of the model parameters. Also, they did not provided an
estimator of the standard error of their model parameter estimator
either.

\cite{Rizoiu2022} proposed a quasi-likelihood type method to estimate
the Hawkes process from interval censored data or counts of events in
different intervals. Their method works by defining the (quasi-)
likelihood by pretending the Hawkes process is an inhomogeneous
Poisson process with an intensity function equal to the mean intensity
process of the Hawkes process and then maximising the thus defined
likelihood as a function of the model parameters. Their method does
not apply to stationary Hawkes processes since the corresponding
Poisson process has only a single parameter, and so the corresponding
quasi likelihood can not be used to distinguish different parameters
of the Hawkes process. They did not address the question of standard
error estimation either.

In this work we propose to approximate the ML estimator (MLE) of the
parameters of the Hawkes process from a discretely observed sample
path using Markov Chain Monte Carlo (MCMC). We first note that MLE is
the maximum a posteriori (MAP) estimator with a flat prior for the
parameter and therefore a posterior density proportional to the
likelihood function. When the amount of data is large, by Laplace's
approximation \citep[see e.g.][$\S$10.2]{vandervaart07}, the posterior
distribution can typically be approximated by a Gaussian distribution
with the mean equal to the ML/MAP estimator and the variance equal to
the inverted negative Hessian matrix of the log-likelihood
function. Therefore, we can approximate the MLE and the confidence
intervals/regions using the posterior mean/median and the credible
intervals/regions respectively. Even when the likelihood function is
available analytically and easy to evaluate, finding the posterior
mean/median and credible intervals analytically can still be difficult in
practice 
and require numerical methods such as the Markov chain Monte Carlo
(MCMC), which constructs an ergodic Markov chain with the desired
posterior distribution as its stationary distribution by e.g. the
Metropolis-Hastings algorithm \citep{Metropolis1953,Hastings1970} and
then uses the occupation measure based on a sufficiently long run of
it to approximate the posterior distribution.

In the context of estimating Hawkes process parameters from a
discretely observed sample path, a further complication is that the
likelihood function is not available analytically. However, we are
able to construct an unbiased estimator of the intractable likelihood
function, which allows us to use the Pseudo-Marginal
Metropolis-Hastings \citep[PMMH][]{Andrieu2009} algorithm to construct
a Markov chain with the correct stationary distribution. Our estimator
of the likelihood is based on the sequential Monte Carlo (SMC), aka
particle filtering method
\citep{GordonEtAl1993,Kitagawa1996}. Therefore, our method might also
be considered a special case of the particle marginal
Metropolis-Hastings sampler of \cite{Andrieu2010}, although they
assumed a hidden Markov model where the hidden state evolves according
to a Markov process and the observations of the state at different
time points are conditionally independent, while in our case the
hidden state process is not Markovian in general and the observations
are not conditionally independent either. Although computationally
intensive, our method is able to produce estimators that are
statistically more efficient than the existing methods, and moreover,
confidence intervals for the parameters are easily available.

The rest of the article is organized as
follows. Section~\ref{sec:datMethod} presents the estimation problem
and the proposed estimation method. Section~\ref{sec:simulations}
assesses the performance of the proposed SMC likelihood estimator and
the approximate ML estimator on simulated data and compare performance
with the spectral method of \cite{Cheysson2022} and the MCEM method of
\cite{Shlomovich2022jcgs}. In Section~\ref{sec:application} we analyse
a weekly measles case count dataset from Tokyo Japan using the
proposed method and compare the results with the previous analysis by
\cite{Cheysson2022}. Section~\ref{sec:discussion} concludes with a
discussion.

\section{The data, the model, and the estimation method}
\label{sec:datMethod}
Let $N(t), t\geq 0$ be the counting process corresponding to a Hawkes
process, with $N(t)$ denoting the number of events of interest in the
interval $(0,t]$. Suppose $N(t)$ is only observed at the the discrete
time points, $0=t_0<t_1<t_2<\dots<t_m$.  We assume that the
observation time points are independent of the process $N(t)$, so they
carries no information on the process and therefore can be treated as
fixed in the inference. In addition to the observation times, the only
data available is the values of process at the observation times,
$N(t_1), \dotsc, N(t_m)$. We shall use the convention $N(0)=0$, so the
data is equivalent to the numbers of events in successive intervals
$(t_{i-1},t_i]$,
$n_1=N(t_1),\ n_2=N(t_2)-N(t_1),\ \dotsc,\ n_m=N(t_m)-N(t_{m-1})$.

Let the background event rate of the Hawkes process $N(t)$ be denoted
by $\nu(t)$, and the excitation kernel function by $g(t)$, so that the
intensity process of $N(t)$ relative to the natural filtration
$\set{\mathcal{F}_t}_{t\geq 0}$, with
$\mathcal{F}_t=\sigma\set{N(s);\ s\leq t}$, is given by
\begin{align}\label{eq:lambda}
  \lambda(t)=\nu(t)+\int_{0}^{t-} g(t-s)\rmd N(s)=\nu(t)+\sum_{i:\tau_i<t} g(t-\tau_i), 
\end{align}
where $\tau_1<\tau_2<\dotsb$ denote the event times
, so that $N(t)=\max\set{i:\ \tau_i\leq t}$. In the parametric
inference problem to be considered in this work, the excitation kernel
function is assumed to take some parametric form with parameters
$\theta_g$, such as the exponential kernel
$g(t;\theta_g)=(\eta /\beta) e^{-t/\beta}, \ t > 0$ with parameters
$\theta_g=(\eta,\beta)$ such that $0\leq \eta<1, \beta >0 $, or the
gamma kernel
$g(t;\theta_g)=\eta \Gamma(\alpha)^{-1} \beta^{-\alpha}
t^{\alpha-1}e^{-t/\beta}, \ t>0$, with parameters
$\theta_g=(\eta,\alpha,\beta)$ such that $0\leq \eta<1$,
$\alpha>0,\ \beta>0$
. Note that, in these parametrisations, the integral of the kernel
function is always $\eta$ and is assumed to be strictly less than 1,
to ensure the asymptotic stationarity of the Hawkes process. The
inference problem we consider is to assume a parametric form for $g$,
and estimate the parameters of the model
$\theta=(\theta_\nu,\theta_g)$ with a discretely observed sample path.

\subsection{Likelihood of the model}
\label{sec:lik}

The likelihood of the Hawkes process relative to a continuously
observed sample path up to some censoring time $T$,
$\set{N(t), 0\leq t\leq T}$, or equivalently, the total number of
events $N=N(0,T]$ up to time $T$ and the exact event times
$\tau_1,\dotsc,\tau_N$, is given explicitly by \citep{Daley2003}
\begin{align*}
  L_{\mathrm{ful}}(\theta)=\braces{\prod_{i=1}^N \lambda(\tau_i) }\exp(-\int_0^T \lambda(t)\rmd t),
\end{align*}
with $\lambda(t)$ given in \eqref{eq:lambda} depending on the
parameters $\theta$. The likelihood in this form can be exactly
evaluated and numerically maximised as a function of $\theta$ to
obtain the maximum likelihood estimator (MLE) of $\theta$.

With observations of the sample path at discrete time points only, the
likelihood of the Hawkes model can be formally written as
\begin{align}\label{eq:lik-inc}
  L_{\mathrm{dis}}(\theta)=\Pr(N((t_{i-1},t_i])=n_i, \ i=1,\dotsc,m).
\end{align}
However, there is no known explicit expressions for the likelihood
function, so obtaining the MLE of the Hawkes process with discrete
observations of the sample path only remains a challenge. In this
work, we propose to obtain the MLE via sequential Monte Carlo (SMC)
approximations to the likelihood functions
.
\subsection{Sequential Monte Carlo approximation to the likelihood}
\label{sec:smc_app}
To present the SMC approximate of the likelihood $L_{\mathrm{dis}}(\theta)$
in~\eqref{eq:lik-inc}, we first rewrite it in the following form
\begin{align}
  L_{\mathrm{dis}}(\theta)&{}=\prod_{i=1}^m p_\theta(n_i|n_{1:i-1} )\label{eq:likdis}\\
                 &{}=\prod_{i=1}^m  \int p_{\theta}(n_i|n_{1:i-1},\tau_{1:N_i})
                   P_\theta(\rmd \tau_{1:N_i}|n_{1:i-1}),  \label{eq:likdis2}
\end{align}
where $n_{1:i}$ is shorthand notation for $(n_1,\dotsc,n_i)$,
$N_i=\sum_{j=1}^i n_j$, and $p_{\theta}(n_i|n_{1:i-1}) $ is shorthand for
$\Pr_\theta\braces{N((t_{i-1},t_i])=n_i|N((t_{i-2},t_{i-1}])=n_{i-1},\dotsc,N((t_{0},t_1])=n_1)}$. Similarly,
$p_{\theta}(n_i|n_{1:i-1},\tau_{1:N_i})$ denotes the conditional
probability of $N((t_{i-1},t_i])=n_i$ given the events
$N((t_{i-2},t_{i-1}])=n_{i-1}$, $\dotsc$, $N((t_{0},t_1])=n_1$ and the
times of all the first $N_i$ events, $\tau_{1:N_i}$, and
$ P_\theta(\rmd \tau_{1:N_i}|n_{1:i-1})$ denotes the conditional
distribution of $\tau_{1:N_i}$ given the numbers of events in the
first $i-1$ observation intervals%
. In the rest of this paper, we drop $\theta$ from the subscripts in
various notations, while their dependence on $\theta$ is silently
understood.

To approximate the integrals in \eqref{eq:likdis2} using Monte Carlo,
we need to generate samples (aka particles) from the predictive
distribution of the hidden event times
$P(\rmd \tau_{1:N_i}|n_{1:i-1})$ and use the empirical distributions
of the particles as an approximation of the predictive distribution,
or from a suitable proposal distribution
that dominates the predictive distribution and then use a weighted
empirical distribution of the particles. Since we need to do this for
all $i=1,\dotsc,m$, we do it in a sequential fashion by reusing the
particles generated in the previous time step. That is, we use the
sequential Monte Carlo method \citep{GordonEtAl1993,Kitagawa1996}.

Take an \iid\ random sample (particles)
$\tau_{1:N_1,1}^{(j)}, \ j=1\dotsc,J$ from a proposal distribution
$Q(\rmd \tau_{1:N_1}|n_1)$ that might depend on $n_1$. Then the
associated importance sampling Monte Carlo (MC) approximation of the
predictive distribution $P(\rmd \tau_{1:N_1})$ is given by
\begin{align*}
  \hat P(\rmd \tau_{1:N_1})=\frac 1 J \sum_{j=1}^J w_1^{(j)}
  \delta_{\tau_{1:N_1,1}^{(j)}} (\rmd \tau_{1:N_1}),  
\end{align*}
with $\delta_x(\cdot)$ denoting the Dirac measure at $x$ and the
importance weights $w_1^{(1:J)}$ given by the values of the
Radon-Nikodym derivative of $P(\rmd \tau_{1:N_1})$ relative to
$Q(\rmd \tau_{1:N_1}|n_1)$ evaluated at the particles,
\begin{align*}
  w_1^{(j)}=\bigg[\frac{P(\rmd \tau_{1:N_1})}{Q(\rmd \tau_{1:N_1}|n_1)}\bigg]_{\tau_{1:N_1}=\tau_{1:N_1,1}^{(j)}}.
\end{align*}
So, the Monte Carlo approximation of $p(n_1)=\int
p(n_1|\tau_{1:N_1})P(\rmd \tau_{1:N_1})$ is given by
\begin{align*}
  \hat p(n_1)=\int
  p(n_1|\tau_{1:N_1})\hat P(\rmd \tau_{1:N_1})= \frac 1 J \sum_{j=1}^J
  w_1^{(j)} p(n_1|\tau_{1:N_1,1}^{(j)} ).
\end{align*}

Suppose we have weighted particle approximations to the predictive
distribution and the likelihood contribution respectively in the
previous time step,
\begin{align*}
  \hat P(\rmd \tau_{1:N_{i-1}}|n_{1:i-2})
  &{}=\frac 1 J \sum_{j=1}^J
    w_{i-1}^{(j)}\delta_{\tau_{1:N_{i-1}, i-1}^{(j)}}(\rmd
    \tau_{1:N_{i-1}}),\\
  \hat p(n_{i-1}|n_{1:i-2})
  &{}=\frac 1 J \sum_{j=1}^J w_{i-1}^{(j)}
    p(n_{i-1}|\tau_{1:N_{i-1}, i-1}^{(j)},n_{1:i-2}).
\end{align*}
Let $Q(\rmd \tau_{N_{i-1}+1:N_i}|n_{1:i},\tau_{1:N_{i-1}})$ be a
proposal distribution for $\tau_{N_{i-1}+1:N_i}$ that dominates its
conditional distribution
$P(\rmd \tau_{N_{i-1}+1:N_i}|n_{1:i-1},\tau_{1:N_{i-1}})$. Note that
similar to \cite{PittShephard1999}, we allow the proposal distribution
to depend on the observation in the $i$th interval, $n_i$. By the
factorisation of the predictive distribution
\begin{align}
  P(\rmd \tau_{1:N_i}|n_{1:i-1})&= P(\rmd
                                  \tau_{1:N_{i-1}}|n_{1:i-1}) P(\rmd
                                  \tau_{N_{i-1}+1:N_i}|n_{1:i-1},\tau_{1:N_{i-1}}) \\
                                &= \frac{p(n_{i-1}|\tau_{1:N_{i-1}},n_{1:i-2})}{p(n_{i-1}|n_{1:i-2})}
                                  P(\rmd\tau_{1:N_{i-1}}|n_{1:i-2})\notag\\
                                &\phantom{=}\times \brackets{ \frac{P(\rmd
                                  \tau_{N_{i-1}+1:N_i}|n_{1:i-1},\tau_{1:N_{i-1}})}{Q(\rmd
                                  \tau_{N_{i-1}+1:N_i}|n_{1:i},\tau_{1:N_{i-1}})}
                                  } Q(\rmd
                                  \tau_{N_{i-1}+1:N_i}|n_{1:i},\tau_{1:N_{i-1}}) 
                                  , \label{eq:factorisation}
\end{align}
we can first sample \(\tau_{1:N_{i-1},i}^{(1:J)}\) from the plug-in
estimate of the \emph{filtering} distribution in the previous time step
\begin{align}
  \hat P(\rmd \tau_{1:N_{i-1}}|n_{1:i-1})& =
                                           \frac{p(n_{i-1}|\tau_{1:N_{i-1}},n_{1:i-2})}{\hat
                                           p(n_{i-1}|n_{1:i-2})}  
                                           \hat P(\rmd
                                           \tau_{1:N_{i-1}}|n_{1:i-2})\notag\\ 
                                         &=
                                           \frac{\frac 1 J \sum_{j=1}^J
                                           p(n_{i-1}|\tau_{1:N_{i-1},i-1}^{(j)},n_{1:i-2}) w_{i-1}^{(j)}
                                           \delta_{\tau_{1:N_{i-1},i-1}^{(j)}}(\rmd
                                           \tau_{1:N_{i-1}})}{\hat
                                           p(n_{i-1}|n_{1:i-2})}. \label{eq:filteringApprox}
\end{align}
Next, for each \(\tau_{1:N_{i-1},i}^{(j)}\), \(j=1,\dotsc,J\),
generate \(\tau_{N_{i-1}+1: N_i, i}^{(j)}\) from the proposal
distribution
\(Q(\rmd
\tau_{N_{i-1}+1:N_i}|\tau_{1:N_{i-1},i}^{(j)},n_{1:i})\). Finally,
define
$\tau_{1:N_i,i}^{(j)}=(\tau_{1:N_{i-1},i}^{(j)},
\tau_{N_{i-1}+1:N_i,i}^{(j)})$, $j=1,\dotsc J$ to be the particles in
the current time step, with respective importance weights
\begin{align}
  w_i^{(j)}=\brackets{ \frac{P(\rmd
  \tau_{N_{i-1}+1:N_i}|n_{1:i-1},\tau_{1:N_{i-1},i}^{(j)})}{Q(\rmd
  \tau_{N_{i-1}+1:N_i}|n_{1:i},\tau_{1:N_{i-1},i}^{(j)})}}_{
  \tau_{N_{i-1}+1: N_{i}}=\tau_{N_{i-1}+1: N_{i}, i}^{(j)}}. \label{eq:filteringWeights}
\end{align}
Then a natural weighted particle approximation to the predictive
distribution in the current time step is given by
\begin{align*}
\hat P(\rmd \tau_{1:N_i}|n_{1:i-1})=\frac1J\sum_{j=1}^J w_i^{(j)}
  \delta_{\tau_{1:N_i, i}^{(j)}}(\rmd \tau_{1:N_i}),
\end{align*}
and the likelihood contributions, or the integrals
in~\eqref{eq:likdis2}, can be approximated by
\begin{align*}
  \hat p(n_i|n_{1:i-1})=\frac 1 J \sum_{j=1}^J w_i^{(j)}
  p(n_i|\tau_{1:N_i, i}^{(j)},n_{1:i-1}), 
\end{align*}
where 
\begin{align}
  p(n_i|\tau_{1:N_i,i}^{(j)},n_{1:i-1})=
  \begin{cases}
    0, & \tau_{N_i}> t_i;\\
    \exp\braces{-\int_{t_{i-1}\vee
    \tau_{N_i,i}^{(j)}}^{t_i}\lambda(s;\tau_{1:N_i, i }^{(j)}) \rmd s}.
       &
         \tau_{N_i}\leq t_i,
  \end{cases}\label{eq:condProb}
\end{align}
Here, the binary operator $\vee$ is defined by
\(x\vee y\stackrel{\mathrm{def}}=\max\set{x,y}\) and has higher
precedence than plus or minus ($\pm$), and
\(\lambda(s;\tau_{1:N_i, i}^{(j)})\) is as in~\eqref{eq:lambda},
except that the event times are replaced by \(\tau_{1:N_i,
  i}^{(j)}\). To evaluate the integral in~\eqref{eq:condProb}, it is
useful to note
\begin{align*}
  \int_{t_{i-1}\vee \tau_{N_i}}^{t_i}\lambda(s;\tau_{1:N_i}) \rmd s =
  \int_{t_{i-1}\vee \tau_{N_i}}^{t_i}\nu (s)\rmd s +
  \sum_{k=1}^{N_i} \braces{G(t_i-\tau_k) -G(t_{i-1}\vee \tau_{N_i} -\tau_{k})},  
\end{align*}
with \(G(\cdot)=\int_0^\cdot g(t)\rmd t\).
Finally, the SMC approximation of the likelihood~\eqref{eq:likdis} is
given by
\begin{align}
  \hat L_{\mathrm{dis}}(\theta)=\prod_{i=1}^m \hat
  p(n_i|n_{1:i-1}). \label{eq:likdisApprox}
\end{align}

In the above SMC procedure, generating \(\tau_{1:N_{i-1},i}^{(1:J)}\)
from the approximate filtering distribution%
~\eqref{eq:filteringApprox} amounts to bootstrap resampling
$\set{\tau_{1:N_{i-1},i-1}^{(j)},\ j=1,\dotsc,J}$ with appropriate
weights. Therefore, the method is also known as bootstrap particle
filtering. The following result says that the bootstrap particle
filter estimate for the likelihood is unbiased. An elementary proof
of it is given in the Appendix \citep[cf. also][]{Andrieu2010,Pitt2012}.
\begin{prop}\label{prop:unbiasedness}
  Viewed as a random variable, where the randomness comes not from the
  data generating process but from the Monte Carlo procedure, the
  bootstrap particle filter approximation given
  in~\eqref{eq:likdisApprox} has an expectation equal to the
  likelihood function~\eqref{eq:likdis}.
\end{prop}

The actual implementation of the SMC procedure described above
requires a suitable choice of the proposal distribution
$Q(\rmd \tau_{N_{i-1}+1:N_i}|\tau_{1:N_{i-1}}, n_{1:i})$.  We choose
the proposal so that $\tau_{N_{i-1}+1:N_i}$ equals in distribution to
the (ordered) times of the first $n_i$ events of a Poisson process on
$(t_{i-1},\infty)$ with rate parameter
$\rho=\gamma_{0.95;n_i,1}/(t_i-t_{i-1})$, where $\gamma_{p;d,r}$
denotes the $p$-quantile of the gamma distribution with $d$ degrees
of freedom and rate parameter $r$. Here the choice of $\rho$ is to
ensure that with a high probability of 0.95 that the $n_i$-th event of
the Poisson proposal process has happened by time $t_i$. With our
choice of the proposal distribution, the Radon-Nikodym derivative
needed to calculate the importance weights~\eqref{eq:filteringWeights}
are given by
\begin{align}
  \frac{P(\rmd
  \tau_{N_{i-1}+1:N_i}|n_{1:i-1},\tau_{1:N_{i-1}})}{Q(\rmd
  \tau_{N_{i-1}+1:N_i}|n_{1:i},\tau_{1:N_{i-1}})}  =
  \frac{\prod_{k=N_{i-1}+1} ^{N_i} \lambda(\tau_{k}) e^{-\int_{t_{i-1}}^{\tau_{N_i}\vee t_{i-1}}
  \lambda(t)\rmd t}}{\rho^{n_i} e^{-\rho (\tau_{N_i}\vee t_{i-1}-t_{i-1})}}.
  \label{eq:deriv}
\end{align}

\begin{rem}
  If $n_i=0$ in any interval $(t_{i-1},t_i]$, then
  $\tau_{N_{i-1}+1:N_i}$ is empty and there is no need to generate
  from the proposal distribution. Particle generation in such
  intervals only requires the bootstrap resampling step, and the
  relative weights of the particles $w_i^{(j)}$
  in~\eqref{eq:filteringWeights} are all $1$.
\end{rem}

\begin{rem}
  If two or more successive intervals all have no events, then such
  intervals should be collapsed to form a single interval with zero
  event counts and the observation time points reduced and relabelled,
  as a data preprocessing procedure. This can lead to substantial
  gains in computational efficiency when the sample path is frequently
  observed at regularly spaced time points, leading to many zero
  counts.
\end{rem}

\begin{rem}
  If the excitation kernel is an exponential function
  $g(t)=\eta /\beta\, e^{- t/\beta}$, then the SMC likelihood
  estimation procedure can be simplified significantly by taking
  advantage of the Markov property of the accumulated excitation
  effect
  \begin{align*}
    \varepsilon(t)=\lambda(t)-\nu(t)=\sum_{k=1}^{N(t-)}\frac{\eta}{\beta}
    e^{-\frac{t-\tau_k}{\beta}},\ t\geq 0.
  \end{align*}
  Specifically, in the $i$th observation interval $(t_{i-1},t_i]$,
  if $n_i=N_i-N_{i-1}=0$ then
  \begin{align*}
  \varepsilon(t)=\varepsilon(t_{i-1}) e^{-(t-t_{i-1})/\beta}, \ t\in(t_{i-1},t_i].
  \end{align*}
  If $n_i\geq 1$, then we have
  \begin{align*}
    \begin{cases}
      \varepsilon(t)=\varepsilon(t_{i-1}) e^{-(t-t_{i-1})/\beta},\
      \lambda(t)={}\nu(t)+\varepsilon(t), \ \ t\in (t_{i-1},
      \tau_{N_{i-1}+1}],\\ 
      \Pr\braces{\tau_{N_{i-1}+1}\in \rmd t}=\lambda(t) e^{-\int_{t_{i-1}}^t \lambda(s)\rmd s} 1_{(t_{i-1},\infty)} (t) \rmd t, \\
      \varepsilon(\tau_{N_{i-1}+1}+) = \varepsilon(\tau_{N_{i-1}+1}) + \eta/\beta;
    \end{cases}
  \end{align*}
  and for $k=N_{i-1}+2,\dotsc,N_i,$
  \begin{align*}
    \begin{cases}
      \varepsilon(t)=\varepsilon(\tau_{k-1}+)
      e^{-(t-\tau_{k-1})/\beta},\ \lambda(t)=\nu(t)+\varepsilon(t), \ \
      t\in (\tau_{k-1}, \tau_{k}],\\ 
      \Pr\braces{\tau_k\in \rmd t}=\lambda(t) e^{-\int_{\tau_{k-1}}^t \lambda(s)\rmd s} 1_{(\tau_{k-1},\infty)} (t) \rmd t, \\
      \varepsilon(\tau_k+)=\varepsilon(\tau_k)+\eta/\beta;
    \end{cases}
  \end{align*}
  and finally
  \begin{align*}
    \varepsilon(t)=\varepsilon(\tau_{N_i}+) e^{-(t-\tau_{N_i})/\beta}, \ t\in (\tau_{N_i},t_i]. 
  \end{align*}
  Therefore, the calculation of the conditional
  probabilities~\eqref{eq:condProb} and the Radon-Nikodym
  derivative~\eqref{eq:deriv} can be simplified by computing the
  $\varepsilon(\tau_k)$'s recursively and noting
  \begin{align*}
    \int_{t_{i-1}}^{t_{i-1}\vee\tau_{N_i}} \lambda(t)\rmd t={}
    & \int_{t_{i-1}}^{t_{i-1}\vee\tau_{N_i}}\nu(t)\rmd
      t+
      \sum_{k=N_{i-1}+1}^{N_i} \varepsilon((\tau_{k-1}\vee t_{i-1})+)
      \beta (1-e^{-(\tau_k - \tau_{k-1}\vee t_{i-1})/\beta}),
    \\
    \int_{t_{i-1}\vee \tau_{N_i}}^{t_i} \lambda(t)\rmd t ={}
    & \int_{t_{i-1}\vee \tau_{N_i}}^{t_i} \nu(t)\rmd t +
      \varepsilon((t_{i-1}\vee \tau_{N_i})+) \beta (1-e^{-(t_i-t_{i-1}\vee\tau_{N_i})/\beta}). 
  \end{align*}

  The storage requirement of the procedure can also be reduced
  drastically. Instead of $\tau_{1:N_{i}, i}^{(1:J)}$, we can take
  $((\varepsilon(t_{i-1}))_i,\tau_{N_{i-1}+1:N_i, i})^{(1:J)}$ to be the
  particles, and moreover, only $(\varepsilon(t_{i-1}))_i^{(1:J)}$ needs
  to be stored 
  while each $\tau_{N_{i-1}+1:N_i}^{(j)}$ can be generated based on
  $(\varepsilon(t_{i-1}))_i^{(j)}$ and discarded after it is used to calculate
  the conditional probability~\eqref{eq:condProb} and the
  corresponding weight~\eqref{eq:filteringWeights} and $(\varepsilon(t_i))_i^{(j)}$.

\end{rem}

\begin{rem}
  A natural and seemingly good proposal distribution for
  $\tau_{N_{i-1}+1:N_i}$ is the distribution of the first $n_i$ event
  times after $t_{i-1}$ of the Hawkes process itself. However, a
  problem with this proposal is that when the trial parameter is
  highly unlikely relative to the observed data, e.g. when the
  background event rate is too low, then it can happen that all the
  $J$ particles generated in an interval have
  $\tau_{N_i,i}^{(j)} > t_i$, leading to all-zero weights in the
  subsequent bootstrap resampling and breakdown of the likelihood
  approximation procedure. In contrast, our choice ensures that such
  all-zero weights are practically impossible; e.g. with as few as 16
  particles, the probability of all particles having zero weights is
  $0.05^{16}=1.5\times 10^{-21}$. Although less serious, another
  problem is that generating the event times of the Hawkes process is
  computationally more demanding than generating the times of a
  Poisson process, which only requires generating exponential random
  variables and then calculating their cumulative sums.
\end{rem}

\begin{rem}
  Another proposal distribution for $\tau_{N_{i-1}+1:N_i}$ is
  considered by \cite{Shlomovich2022jcgs}, which generates the
  $\tau_k$, $k=N_{i-1}+1, \dotsc, N_i$ sequentially according to the
  intensity of the Hawkes process with truncation by $t_i$. While this
  method ensures that all particles will have positive weights, a
  problem is that the weights of the particles are difficult to
  compute, since the conditional distribution of the first $n_i$ event
  times of a Hawkes process given that they are less than a known
  constant is not available in closed form. Also, our numerical
  experimentation reveals that the events generated by this proposal
  in each observation interval have a strong tendency to pile towards
  the right end of the interval.
\end{rem}

\subsection{Estimation of the parameters}
\label{sec:est}
Assume suitable regularity conditions so that when the number of
observation intervals $m$ is large, the loglikelihood
$\log L(\theta)=\log p_\theta(n_{1:m})$ can be approximated by a
quadratic form, that is,
\begin{align*}
\log L(\theta)\approx \log p_{\hat\theta}(n_{1:m}) - \frac12(\theta-\hat \theta)\tr
(- H(\hat\theta)) (\theta-\hat\theta))
\end{align*}
with $H(\theta)$ being the Hessian matrix of the loglikelihood and
$\hat\theta$ the maximizer of the likelihood function, or the maximum
likelihood estimator (MLE) of the parameter vector $\theta$. Then the
likelihood function itself is approximately proportional to an
unnormalized Gaussian density function
$\exp(- \frac12(\theta-\hat \theta)\tr (- H(\hat\theta))
(\theta-\hat\theta)))$. This means that with suitable regularity
conditions, the distribution function over the parameter space
$\Theta$ with a density relative to the Lebesgue measure proportional
to the likelihood function $L(\theta)$, which we shall refer to as the
likelihood distribution, should be asymptotically Gaussian with mean
equal to the MLE and variance equal to the inverse of the negative
Hessian matrix evaluated at the MLE. Therefore, if we approximate the
likelihood distribution, say by using the Markov Chain Monte Carlo
(MCMC) method, then we can easily obtain an approximation to the MLE
and the Hessian matrix by extracting the mean/median and variance of
the Monte Carlo sample. To obtain approximate Wald confidence
intervals for the parameters, we can simply extract the appropriate
sample quantiles of the corresponding margins of the Monte Carlo
sample. Note that from a Bayesian perspective, the likelihood
distribution is simply the posterior distribution of the parameter
vector when the prior distribution is the (possibly improper) uniform
distribution, and so the proposed approximate MLE and confidence
intervals can also be interpreted as the posterior mean/median and the
credibility intervals in Bayesian inference.

Since the density of the likelihood distribution is only known up to a
constant, a natural method to construct a Markov Chain to generate
samples from it is the Metropolis-Hastings (MH) algorithm, where the
transition kernel is such that given the current state $\theta$ of the
chain, the next state $\theta'$ is a random draw $\theta^*$ from some
prespecified proposal distribution $Q(\rmd \theta^* |\theta)$ that
dominates $L(\theta^*)\rmd \theta^*$ if $\theta^*$ meets a (random)
criterion, or the current state $\theta$ otherwise. The random
acceptance criterion here is that $A(\theta^*,\theta)\geq U$, for a
uniform random number $U$ on the interval [0,1] and
\begin{align}\label{eq:acptProb}
  A(\theta^*,\theta) =  \frac{L(\theta^*)\rmd \theta^*}{Q(\rmd \theta^*|\theta)} \Big/
  \frac{L(\theta)\rmd\theta}{Q(\rmd
  \theta|\theta^*)}=\frac{L(\theta^*)\rmd \theta^* Q(\rmd
  \theta|\theta^*)}{L(\theta)\rmd\theta Q(\rmd \theta^*|\theta)}.   
\end{align}
When the proposal distribution is symmetric in $\theta$ and
$\theta^*$, the acceptance ratio $A(\theta^*,\theta)$ can be
simplified to the likelihood ratio $L(\theta^*)/L(\theta)$.

In the context of Hawkes process with discretely observed data, the
likelihood $L(\theta)=L_{\mathrm{dis}}(\theta)$ is not available, but
we have an unbiased estimator, so we can use the Pseudo-Marginal
Metropolis-Hastings \citep[PMMH][]{Andrieu2009} algorithm to construct
a Markov chain $\theta^{(i)}, \ i=1,2,\dotsc$ where the transition
kernel is the same as in the MH algorithm but the likelihood function
needed in the calculation of the acceptance ratio~\eqref{eq:acptProb}
is replaced by the unbiased estimator
$\hat L(\theta)=\hat L_{\mathrm{dis}}(\theta)$
in~\eqref{eq:likdisApprox}. Note that in calculating the acceptance
ratio $A(\theta^*,\theta^{(i)})$, $\hat L(\theta^{(i)})$ should
\emph{not} be recalculated; rather, its value from before when
$\theta^{(i)}$ was last accepted should be used, although the random
number $U$ to compare $A(\theta^*,\theta^{(i)})$ against should be
freshly generated each time a $\theta^*$ is proposed. It can be shown
that the stationary distribution of the thus constructed Markov chain
is equal to the likelihood distribution, which justifies our use of its
occupation measure to approximate the likelihood distribution. For
completeness, a short proof is given in the Appendix.
\begin{rem}
  The likelihood distribution is a special case of the
  \emph{confidence distribution} discussed e.g. in \cite{Xie2013}. The
  density of the likelihood distribution is known as the normalised
  likelihood function, and its use in statistical inference has been
  discussed in the literature, see
  e.g. \cite{Shcherbinin1987}. 
\end{rem}

\section{Simulation Studies}
\label{sec:simulations}

\subsection{SMC estimator of the likelihood}
\label{sec:likapp}
\begin{figure}[hbt]
  \centering
  \includegraphics[width=0.9\textwidth]{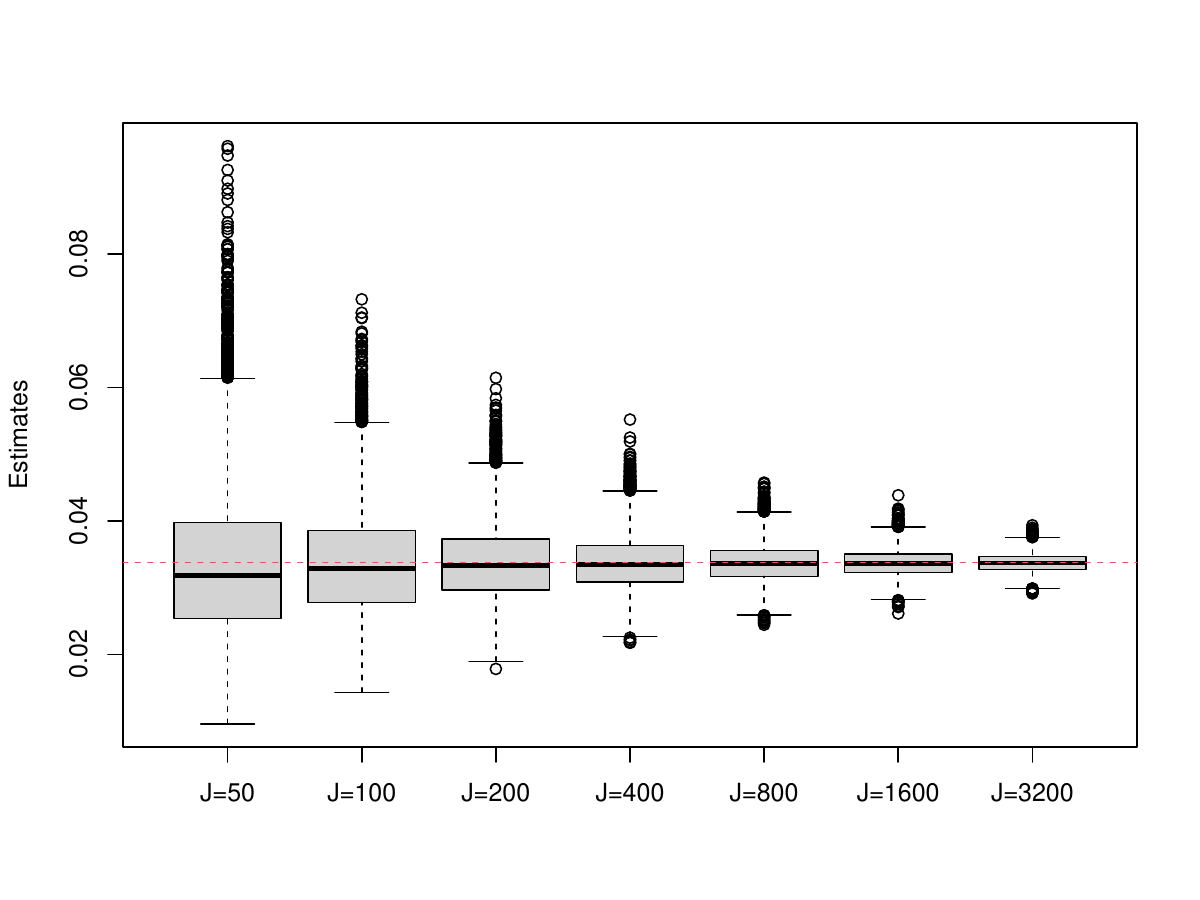}
  \caption[Boxplots of the SMC Estimates of a path probability]{The
    boxplots of 1000 SMC estimates with different $J$ values of the
    probability $\Pr\{N(0,1]=1,N(1,2]=2\}$ for a Hawkes process $N$
    with $\nu=1$, $\eta=0.6$, and a gamma excitation kernel with shape
    parameter 2 and scale parameter 0.1. The dashed horizontal line
    indicates the estimate of the probability via brute force Monte
    Carlo based on 100,000,000 simulated sample paths.}
  \label{fig:prEstBoxplot}
\end{figure}

We present numerical evidence to corroborate the unbiasedness of the
likelihood approximation. For a Hawkes process with constant
background $\nu=1$, branching ratio $\eta=0.6$, and a gamma density
kernel with shape parameter 2 and scale parameter 0.1, the
probability of having $n_1=1$ event in the interval $(0,1]$ and
$n_2=2$ events in the interval $(1,2]$ is around $0.0338$ by
simulating and inspecting 100,000,000 sample paths of the Hawkes
process. The boxplots of 1000 replicates of the SMC approximation of
the probability with different numbers of particles $J$ are shown in
Figure~\ref{fig:prEstBoxplot}, from which we see that the SMC
estimates with different $J$ values are all distributed around the
unbiased brute force Monte Carlo estimate with decreasing variability
when the $J$ value increases, suggesting unbiasedness of the SMC
estimator.
\subsection{Approximate MLE via the Pseudo-Marginal Metropolis-Hastings (PMMH) MCMC}
In this section, we use simulation experiments to investigate the
finite sample performance of the PMMH-MCMC estimator of the model
parameters. We simulate a Hawkes process with constant background
intensity $\nu=2$, branching ratio $\eta=0.6$, and an exponential
density kernel with scale or mean parameter $\beta=0.25$. The
observation time points are evenly distributed between $0$ and $T=100$
or $T=200$ with constant spacing $\Delta=0.1,0.2,0.5$ or $1.0$. The
different $\Delta$ values correspond to different levels of time
coarsening. The number of particles in SMC likelihood approximation is
set at $J=256$. The Hawkes process was simulated 500 times, and for
each simulated sample path, only the values at the times
$0,\Delta,2\Delta,\dotsc,T$ were observed and used in estimating the
parameters $\theta=(\nu,\eta,\beta)$. In implementing the PMMH-MCMC,
we used a Gaussian random walk proposal on the transformed parameters
$(\log\nu,\log(\eta/(1-\eta)),\log\beta)$ with standard deviation
$\sigma=0.05$. The constructed Markov Chain for the transformed
parameter vector was initiated with a random draw from the
3-dimensional standard normal distribution and iterated a total of
50,000 times. For each parameter, the median of the Monte Carlo sample
was taken as the value of the approximate MLE, the lower and upper 2.5
percentile points were taken as the lower and upper limits of the
approximate 95\% Wald confidence interval, and the difference of these
two percentiles divided by $2\Phi^{-1}(0.975;0,1)\approx 3.92$ was
taken as the estimate of the standard error (SE) of the approximate
MLE. For comparison, the case $\Delta=0$ (with exact observations of
event times) is also considered. Note that in this case, the exact
log-likelihood can be computed, and so the MLE and its SE can be
obtained by directly minimizing the negative log-likelihood function
using general purpose numerical optimization routines and inverting
the Hessian matrix. In our implementation, we used the \texttt{optim}
function in \texttt{R} \citep{RCoreTeam2022} for numerical optimization
with the loglikelihood function computed with the aid of the
\texttt{R} package \texttt{IHSEP} \citep{Chen2022a}. The PMMH-MCMC
estimator in the cases with $\Delta>0$ is implemented in
\texttt{julia} \citep{Bezanson2017julia}.

The summary of the 500 estimates for each combination of $\Delta$ and
$T$ are shown in Table~\ref{tab:simEstMCMC}, where (and whereafter)
Est denotes the average of 500 MLE estimates, SE the empirical SE of
the MLE calculated as the standard deviation of the 500 estimates,
$\hat{\mathrm{SE}}$ the average of the SE estimates, and CP the
empirical coverage probability of the 95\% Wald confidence interval
calculated as the proportion of the 500 CIs that contain the
corresponding true parameter value.
\begin{table}[hbt]
  \centering
  \setlength{\tabcolsep}{2pt}
  \begin{tabular}{|cc| cccc| cccc |cccc| }
    \hline
    & & \multicolumn{4}{|c|}{$\nu=2$} &  \multicolumn{4}{|c|}{$\eta=0.6$} &  \multicolumn{4}{|c|}{$\beta=0.25$} \\\cline{3-14}
    T&$\Delta$&Est&SE&$\hat{\mathrm{SE}}$&CP&Est&SE&$\hat{\mathrm{SE}}$&CP&Est&SE&$\hat{\mathrm{SE}}$&CP\\\hline
    100&0  &2.053& 0.2793& 0.2849& 0.954& 0.586& 0.0602& 0.0608& 0.952& 0.248& 0.0494& 0.0472& 0.914\\
    200&0  &2.022& 0.2041& 0.2026& 0.948& 0.596& 0.0421& 0.0430& 0.948& 0.252& 0.0353& 0.0335& 0.952\\\hline
    100&0.1&2.039& 0.3065& 0.2946& 0.942& 0.593& 0.0631& 0.0635& 0.944& 0.256& 0.0501& 0.1104& 0.946\\
    200&0.1&2.005& 0.2047& 0.2041& 0.952& 0.600& 0.0426& 0.0433& 0.960& 0.256& 0.0362& 0.0353& 0.944\\\hline
    100&0.2&2.046& 0.3103& 0.2932& 0.932& 0.591& 0.0639& 0.0631& 0.944& 0.255& 0.0524& 0.0540& 0.944\\
    200&0.2&2.005& 0.2042& 0.2034& 0.938& 0.600& 0.0422& 0.0434& 0.960& 0.255& 0.0378& 0.0363& 0.940\\\hline
    100&0.5&2.054& 0.3222& 0.3036& 0.936& 0.590& 0.0655& 0.0651& 0.954& 0.254& 0.0603& 0.0642& 0.956\\
    200&0.5&2.006& 0.2128& 0.2085& 0.936& 0.600& 0.0445& 0.0444& 0.960& 0.255& 0.0452& 0.0421& 0.934\\\hline
    100&1.0&2.086& 0.3638& 0.3144& 0.910& 0.583& 0.0754& 0.0671& 0.912& 0.242& 0.0952& 0.0834& 0.926\\
    200&1.0&2.014& 0.2341& 0.2108& 0.906& 0.598& 0.0483& 0.0446& 0.924& 0.253& 0.0609& 0.0603& 0.896\\\hline
  \end{tabular}
  \caption{Summary of the parameter estimates in the simulation
    experiments.
  }
  \label{tab:simEstMCMC}
\end{table}
The simulation results suggest that the estimators for all three
parameters have performance similar to the MLE. For example, the
empirical biases of the estimators are negligible compared to their
respective standard errors. When the total observation time $T$
doubles, the standard errors all reduce roughly by a factor of 70\%
($\approx 1/\sqrt2$). We also observe that as the degree of data
coarsening intensifies, the empirical biases and standard errors of
the estimators tend to increase. Meanwhile, the empirical coverage
probabilities of the confidence intervals tend to decrease. This trend
mirrors the growing difficulty the inference problem due to increasing
level of information loss.
\subsection{Comparison with existing methodologies}
\begin{figure}[hbt]
  \centering
  \includegraphics[width=0.495\textwidth]{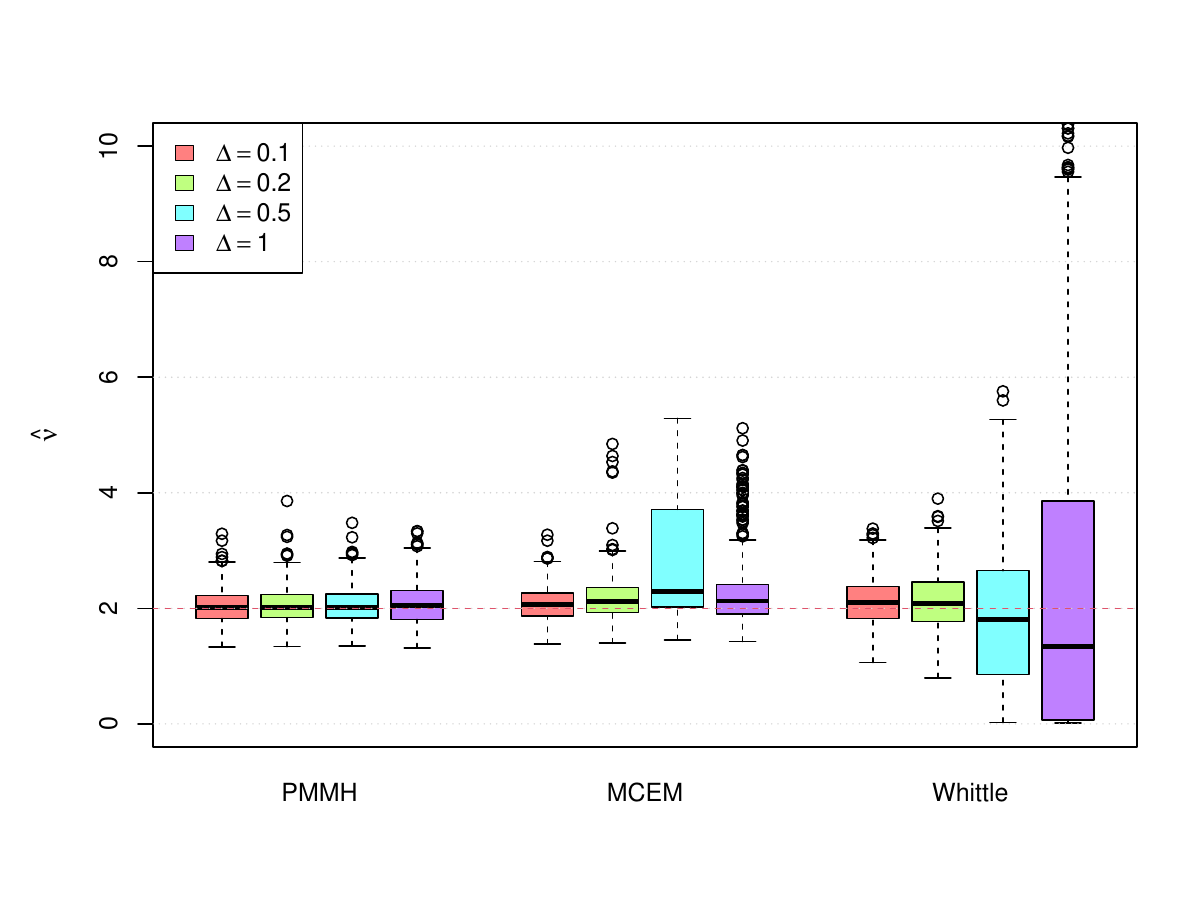} \includegraphics[width=0.495\textwidth]{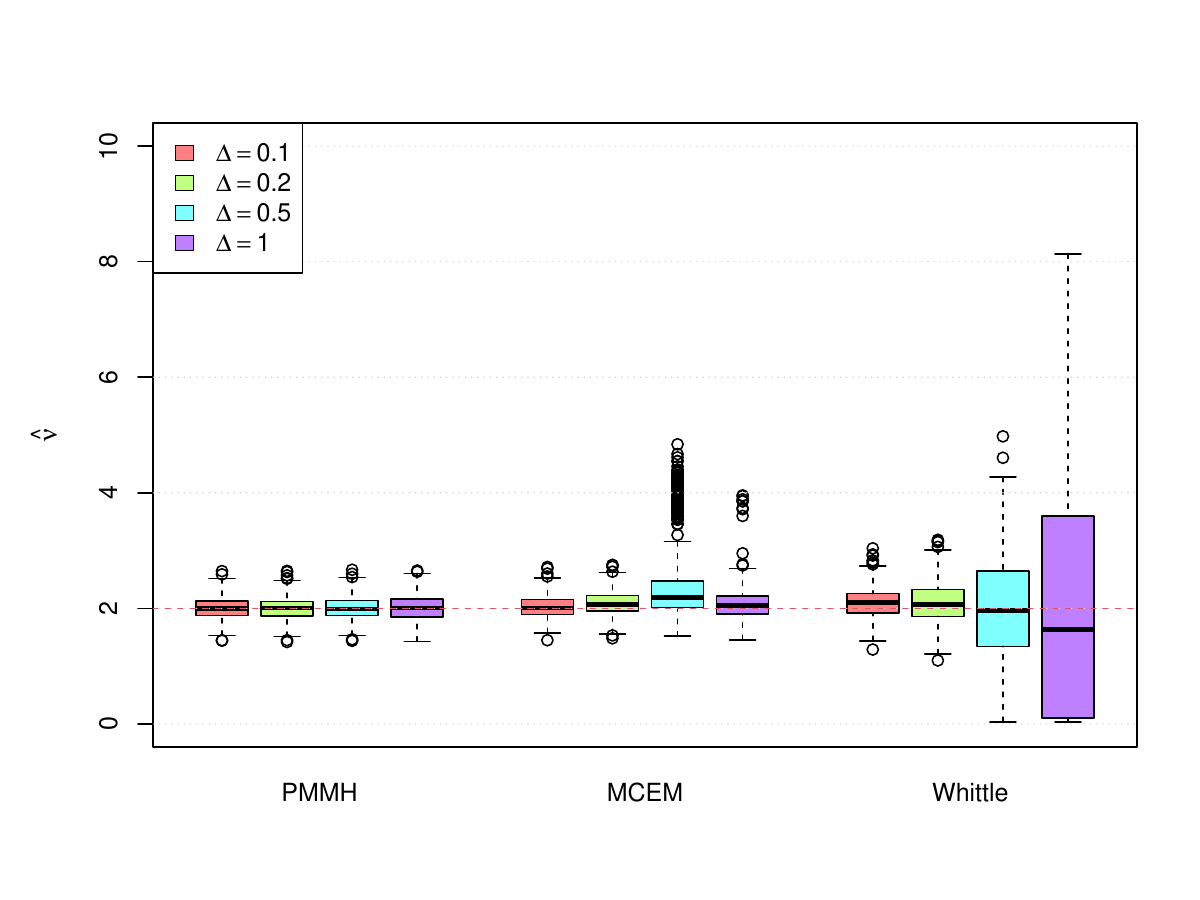}\\
  \includegraphics[width=0.495\textwidth]{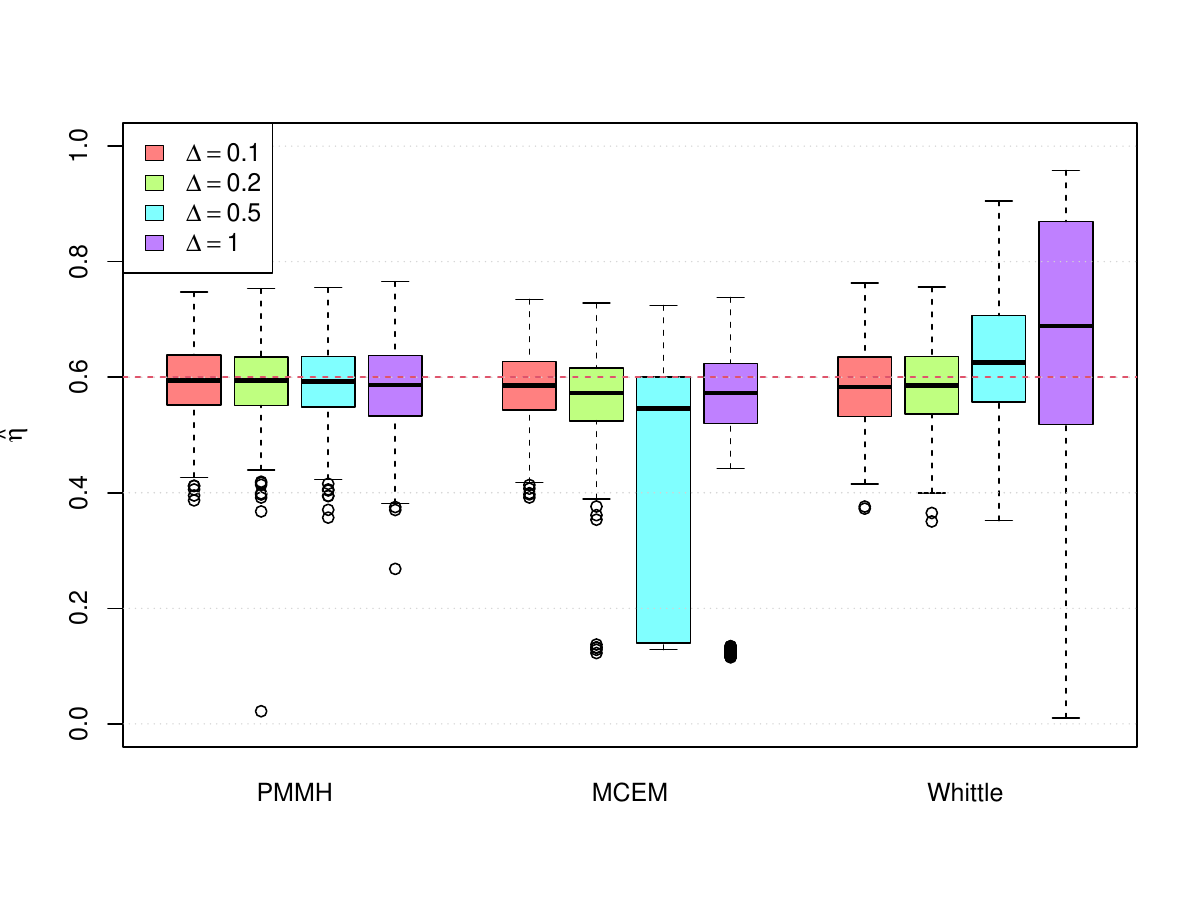} \includegraphics[width=0.495\textwidth]{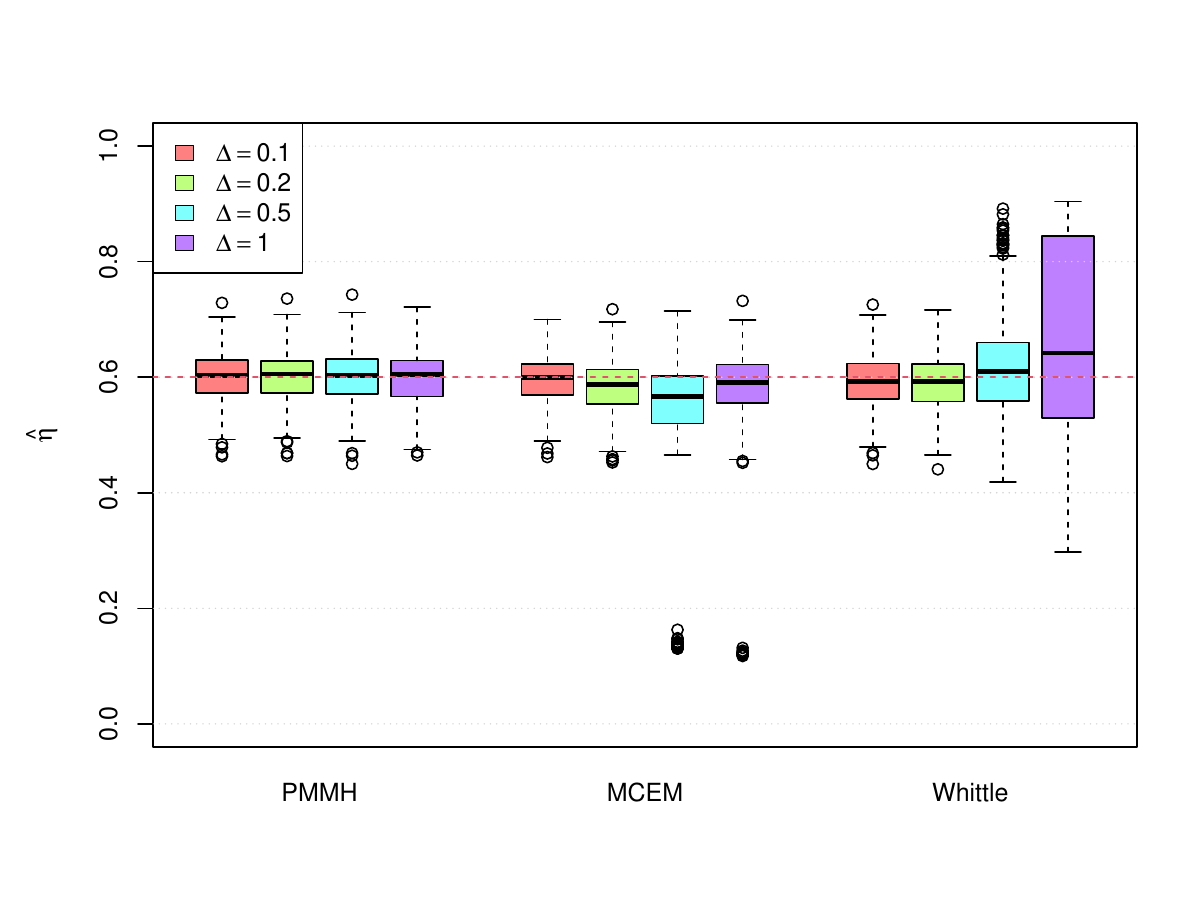}\\
    \includegraphics[width=0.495\textwidth]{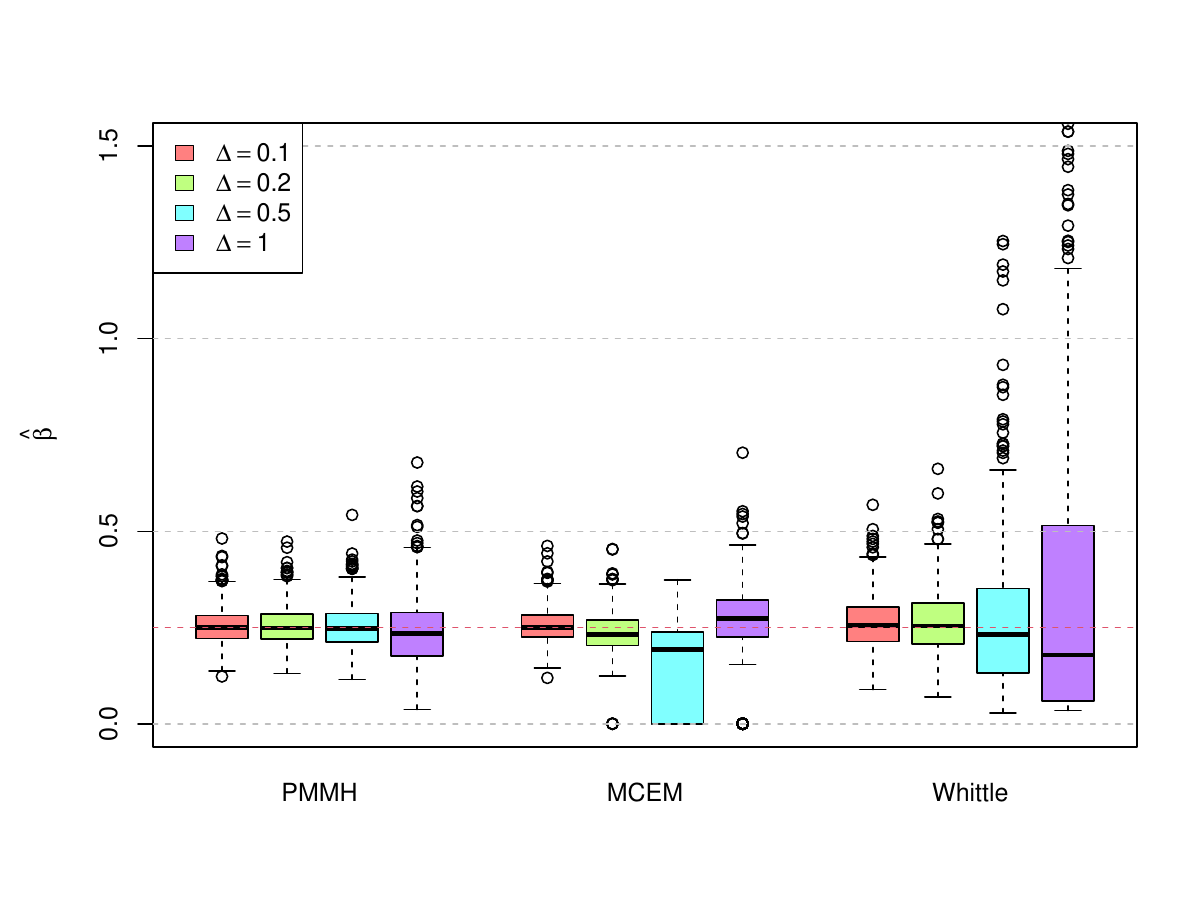} \includegraphics[width=0.495\textwidth]{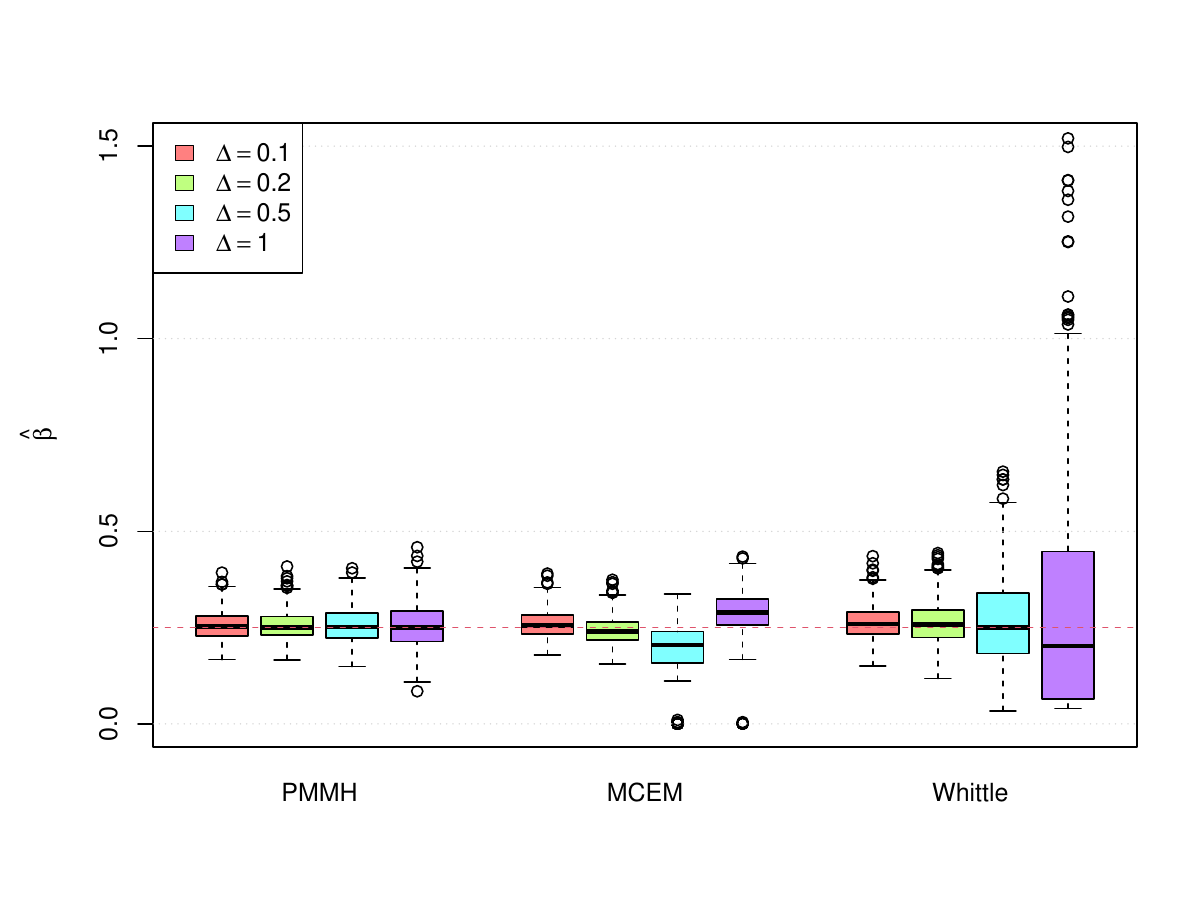}\\
  \caption{Boxplots of the estimates for the parameters of the Hawkes
    process with constant background intensity based on simulated
    data. Left panels: $T=100$, right panels: $T=200$; top panels:
    background intensity $\nu$, middle panels: branching ratio $\eta$,
    lower panels: the scale parameter $\beta$ of the exponential
    excitation kernel. The horizontal dashed lines indicate the true
    values of each parameter ($\nu=2$, $\eta=0.6$, $\beta=0.25$).} 
  \label{fig:simestboxplots}
\end{figure}

In this section we compare the performance of the proposed PMMH
estimator with two competing methods in the literature, namely the
maximum Whittle likelihood method of \cite{Cheysson2022} and the MCEM
(Monte Carlo Expectation Maximization) method of
\cite{Shlomovich2022jcgs}, by applying them on simulated data. The
simulation models are the same as in the previous subsection. The 500
simulated data sets and the PMMH estimates for each of data coarsening
level $\Delta\in\set{0.1,0.2,0.5,1}$ and total observation time
$T\in\set{100,200}$ combination are also the same as before. The MCEM
estimates based on the same simulated data sets for each of $\Delta$
and $T$ combination are computed using the Matlab code released by
\cite{Shlomovich2022jcgs} at this link:
\url{https://github.com/lshlomovich/MCEM-Univariate-Hawkes}. The
Whittle likelihood estimates are computed using the \texttt{whittle}
function in the R package \texttt{hawkesbow} by
\cite{Cheysson2021Rpackage}. The boxplots showing the estimates for
the three parameters by different methods with values of $\Delta$ and
$T$ are shown in Figure~\ref{fig:simestboxplots}.

By Figure~\ref{fig:simestboxplots} it is fairly clear that when the
data coarsening level is low, e.g. when $\Delta=0.1$, all three
estimators have more or less the same performance. However, when
the level of data coarsening increases, the PMMH estimator tends to
perform better than the MCEM and Whittle estimators in terms of both
bias and variance. When $\Delta=0.5$, the body of the boxplot for the
MCEM estimates for each of the parameters seems to lie entirely on one
side of the true value of the corresponding parameter, even with the
larger total observation time, suggesting significant bias of the MCEM
estimator. On the other hand, the Whittle estimator for each of the
three parameters shows rather large variance compared with the other
two estimators, although its bias seems negligible relative to its
standard deviation. The larger variances of the Whittle estimator
suggest that it is less efficient than the likelihood based PMMH
estimator. The bias of the MCEM estimators of
\cite{Shlomovich2022jcgs} might be due to the way they estimated the
Q-function, or the conditional expectation of the complete-data
log-likelihood function given the observed data, in the E-
(Expectation-) step of the EM cycle using Monte Carlo method. Rather
than taking a random sample from the proposal distribution and using
the importance-weighted sample to approximate the conditional
distribution of the hidden state, they calculated the most likely
state(s) according to the proposal distribution and used the
importance-weighted states to approximated the needed conditional
distribution of the hidden state. While it intuitively makes sense,
there is no guarantee that this method leads to unbiased estimate of
the Q-function in the E-step of the EM algorithm, and it seems that
the bias here has been inherited by their final estimator of the model
parameters.

\section{Application}
\label{sec:application}
\begin{figure}[hbt]
  \centering
  \includegraphics[width=0.8\textwidth]{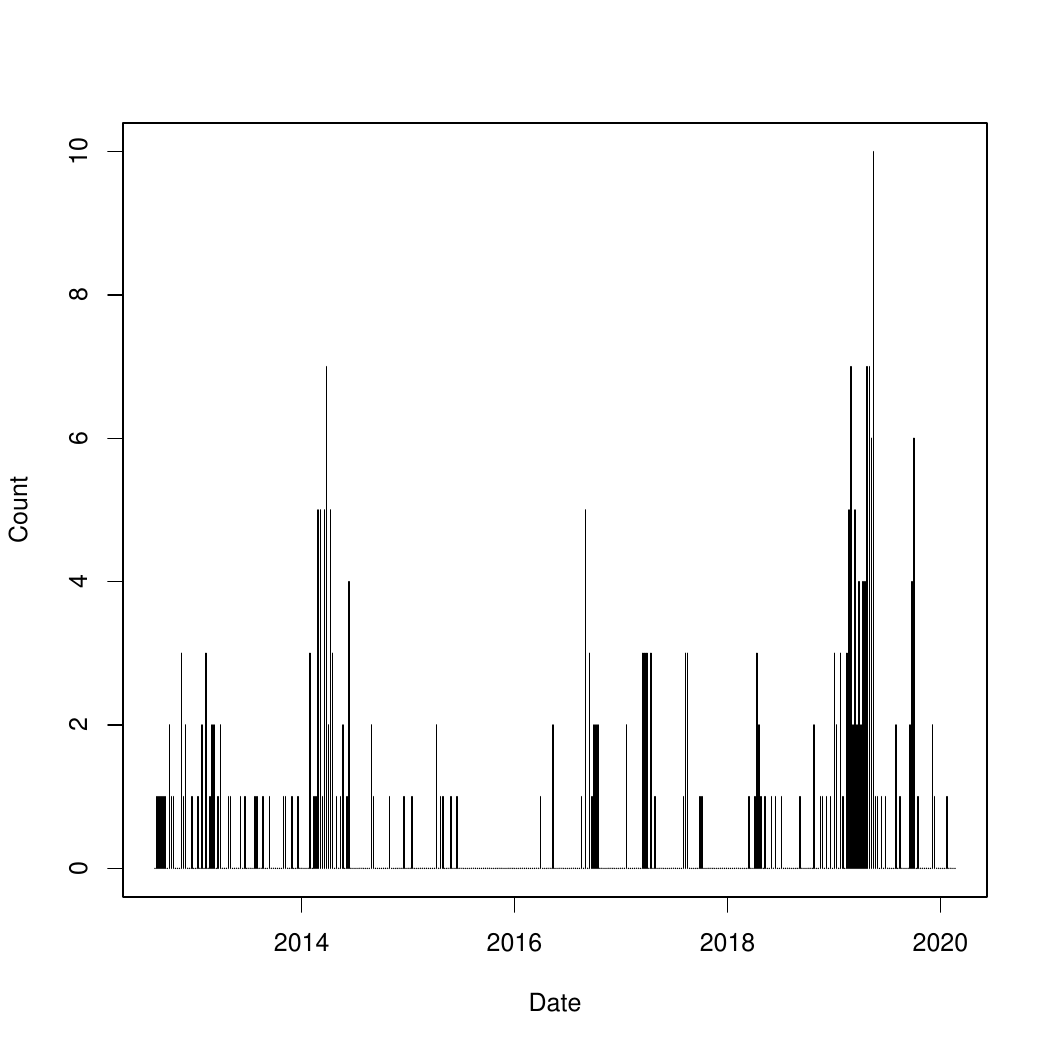}
  \caption[Tokyo measles case counts]{Time series plot of the weekly
    measles case counts in Tokyo Japan from 2012-08-10 to 2020-02-20.}
  \label{fig:tokyomeaslescasecounts}
\end{figure}

In this section we apply the proposed estimation method to an
infectious disease dataset previously studied by
\cite{Cheysson2022}. The dataset contains the weekly counts of measles
cases in Tokyo Japan during the 393-week period from 10 Aug 2012 to 20
Feb 2020. The weekly measles cases counts vary between 0 and 10, with
an average of 0.67, a median of 0.0, and a standard deviation of
1.38. A graph of the weekly case counts against week end date is show
in Figure~\ref{fig:tokyomeaslescasecounts}. \cite{Cheysson2022}
assumed a non-causal Hawkes process with a constant background event
intensity and a Gaussian kernel for the underlying point process of
measles cases and applied the maximum Whittle likelihood estimator to
estimate the model parameters. They estimated a background intensity
of 0.04 day$^{-1}$, a branching ratio of 0.72, and a mean and standard
deviation for the Gaussian excitation kernel 9.8 days and 5.9 days
respectively. They also used formal tests to confirm the fitted
non-causal Hawkes process provides acceptable fit to the data. 

\begin{figure}[hbt]
  \centering
  \includegraphics[width=0.49\textwidth]{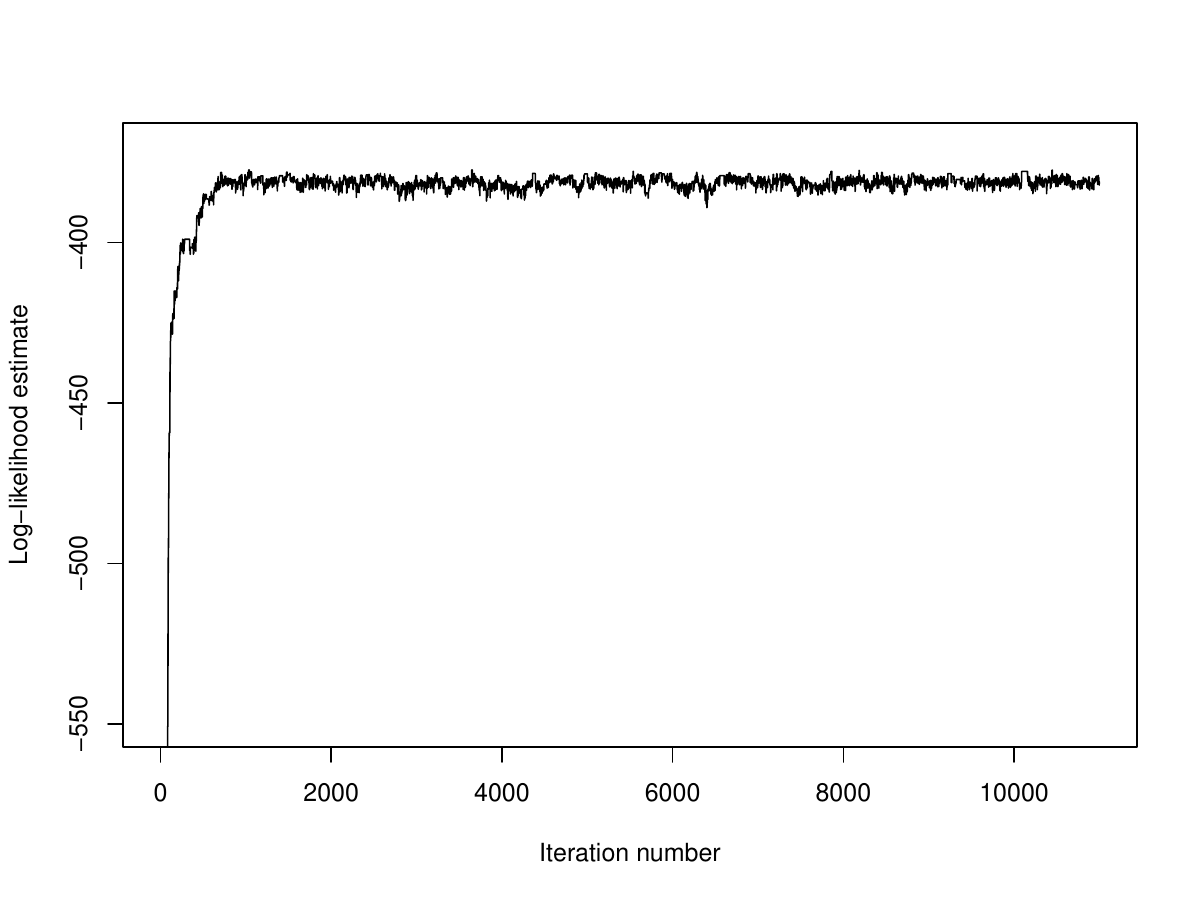}
  \includegraphics[width=0.49\textwidth]{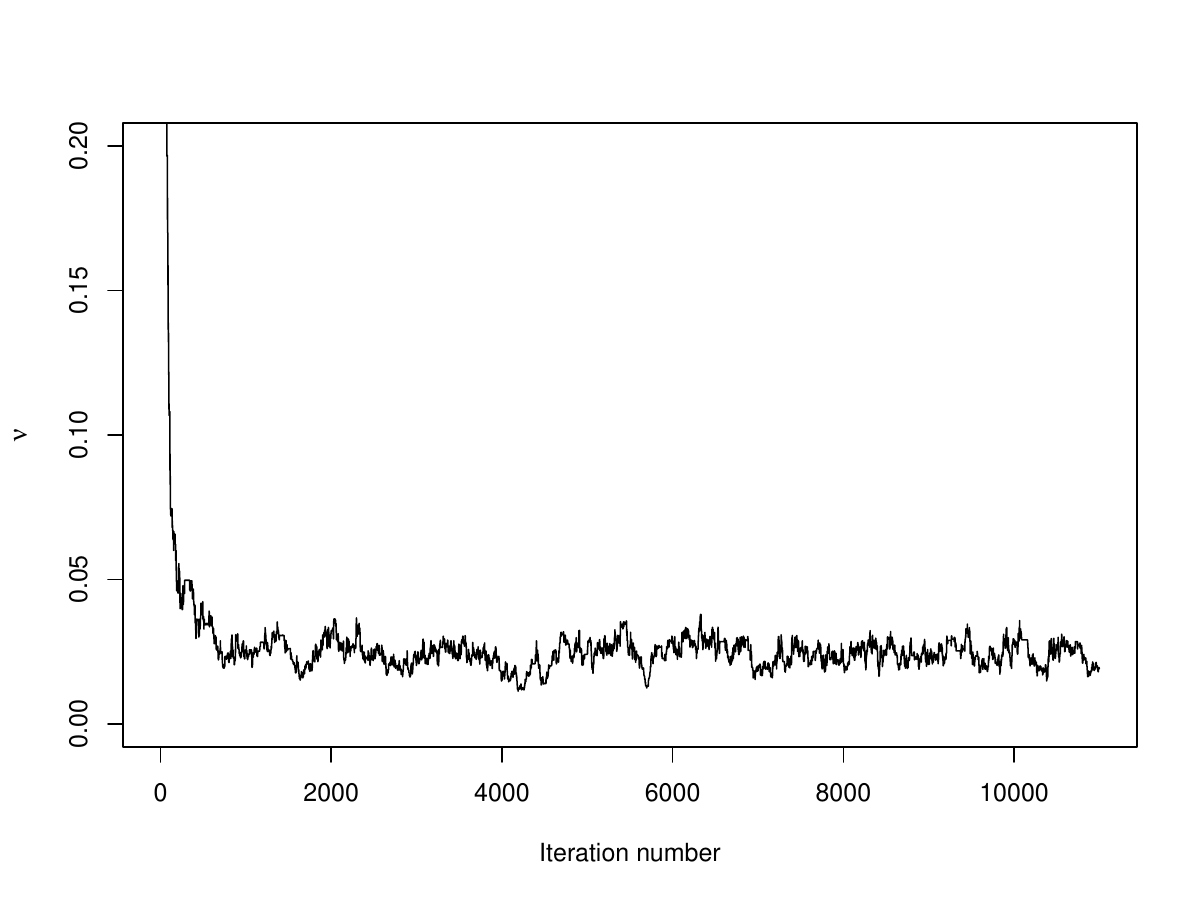}\\
  \includegraphics[width=0.49\textwidth]{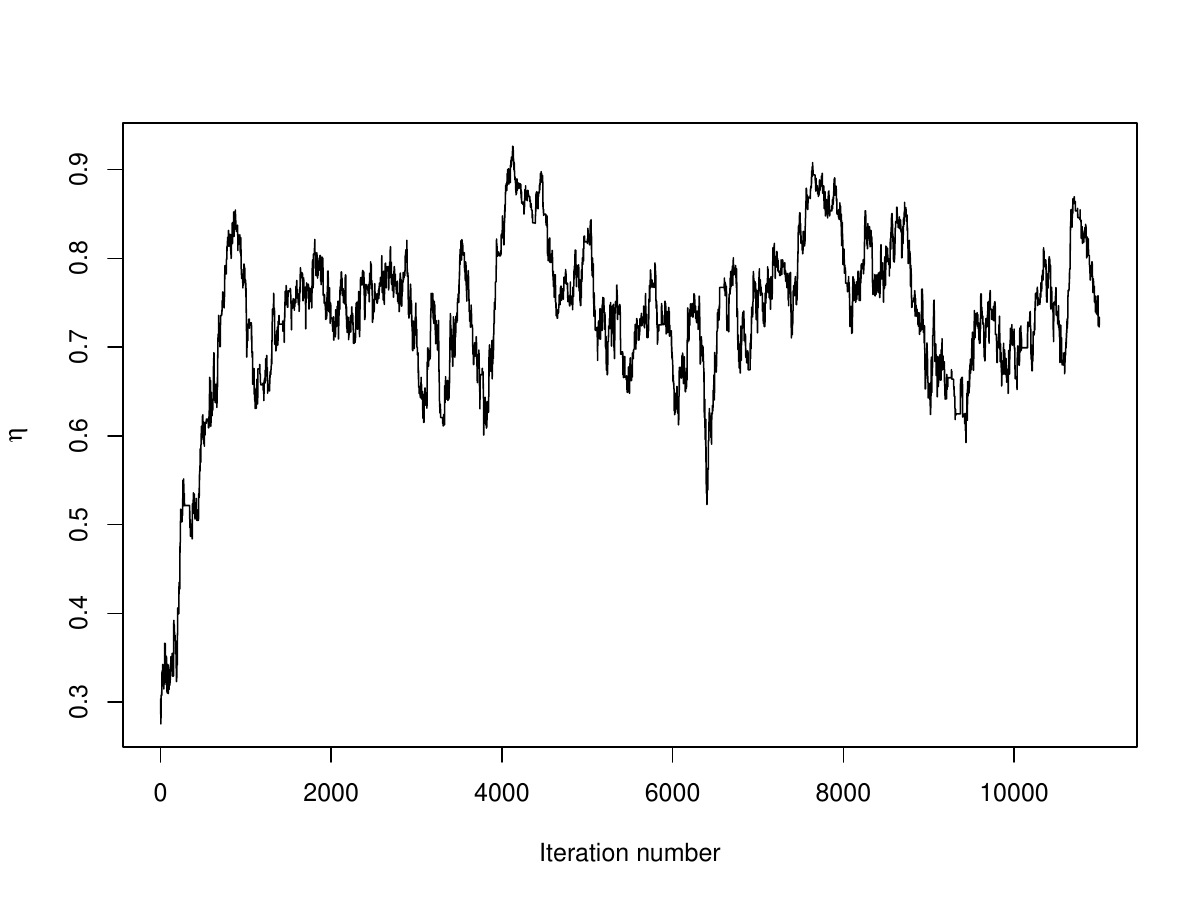}
  \includegraphics[width=0.49\textwidth]{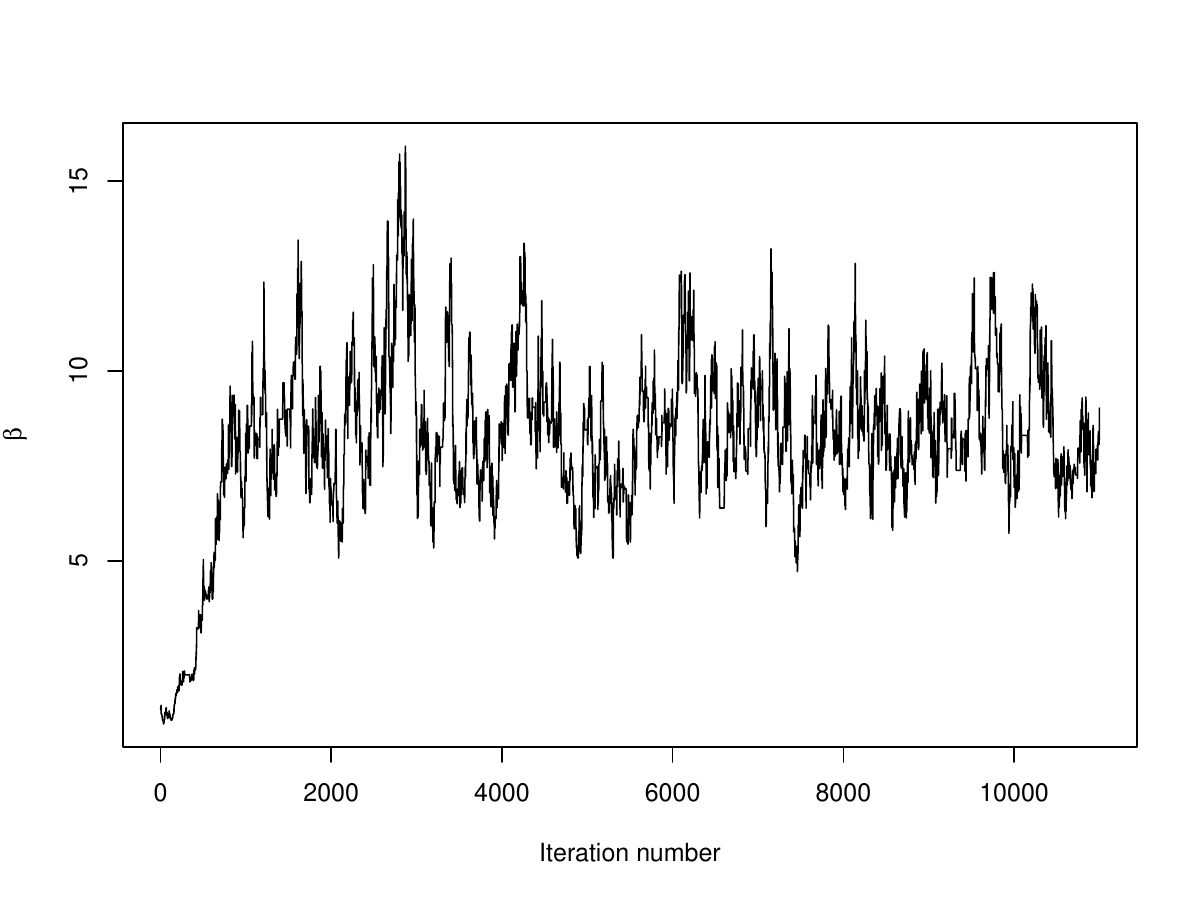}
  \caption[MCMCtrace]{Trace plots of the log-likelihood estimate and the
    parameter values (different dimensions of the MCMC draw).}
  \label{fig:MCMCtraceplot}
\end{figure}

While the fitted non-causal Hawkes process of \cite{Cheysson2022} can
reproduce the spectral features of the data, some of the appealing
features of the classical (causal) Hawkes process is comprised. For
example, the conditional intensity process relatively to the natural
filtration induced by the underlying point process, which is easily
available in the classical Hawkes process, is not tractable
anymore. In this work, we fit a causal Hawkes process with an
exponential kernel function to the same dataset using the PMMH-MCMC
method. We initialise the Markov chain with a random starting point as
in the simulation experiments and iterate the chain for a total of
11000 time steps. Again, we use the random walk Gaussian proposal on
the transformed parameter vector with a standard deviation of 0.05. The
number of particles used in the estimation of the log-likelihood
function is also the same as in the simulation experiments, namely
$J=256$. From the trace plots of the log-likelihood estimate and the
MCMC draw in Figure~\ref{fig:MCMCtraceplot}, it is clear that the
Markov chain has reached convergence after around 1000 iterations. 
Extracting the appropriate quantiles of the next 10,000 MCMC draws, we
obtained the point estimates and 95\% confidence intervals (in
brackets) for the three parameters as follows:
$\hat\nu=0.0244\ ([0.0155,0.0323])$,
$\hat\eta=0.749\ ([0.625,0.882])$, and
$\hat\beta=8.34\ ([5.99,12.45])$. Note that with these parameter
estimates, the expected number of events per 7-day period should be
$\hat\nu/(1-\hat\eta)\times7\approx 0.680$, which is fairly close to the
average weekly case count of $0.672$. In contrast, by the estimates of
\cite{Cheysson2022}, the expected weekly number of cases would be
$0.04/(1-0.72)\times7=1.0$, which is substantially higher than the
corresponding average in the data. The expected number of imported
cases by our model is $\hat\nu\times 393\times7\approx 67$, or 25.4\% of the 264
cases in total, which is slightly lower than the estimated percentage
of $28\%$ by \citeauthor{Cheysson2022}'s non-causal Hawkes process
model, but is closer to the finding of a previous study by
\cite{Nishiura2017}, in which 23 among 103 confirmed cases were found
to be imported cases.

\begin{figure}[hbt]
  \centering
  \includegraphics[width=0.49\textwidth]{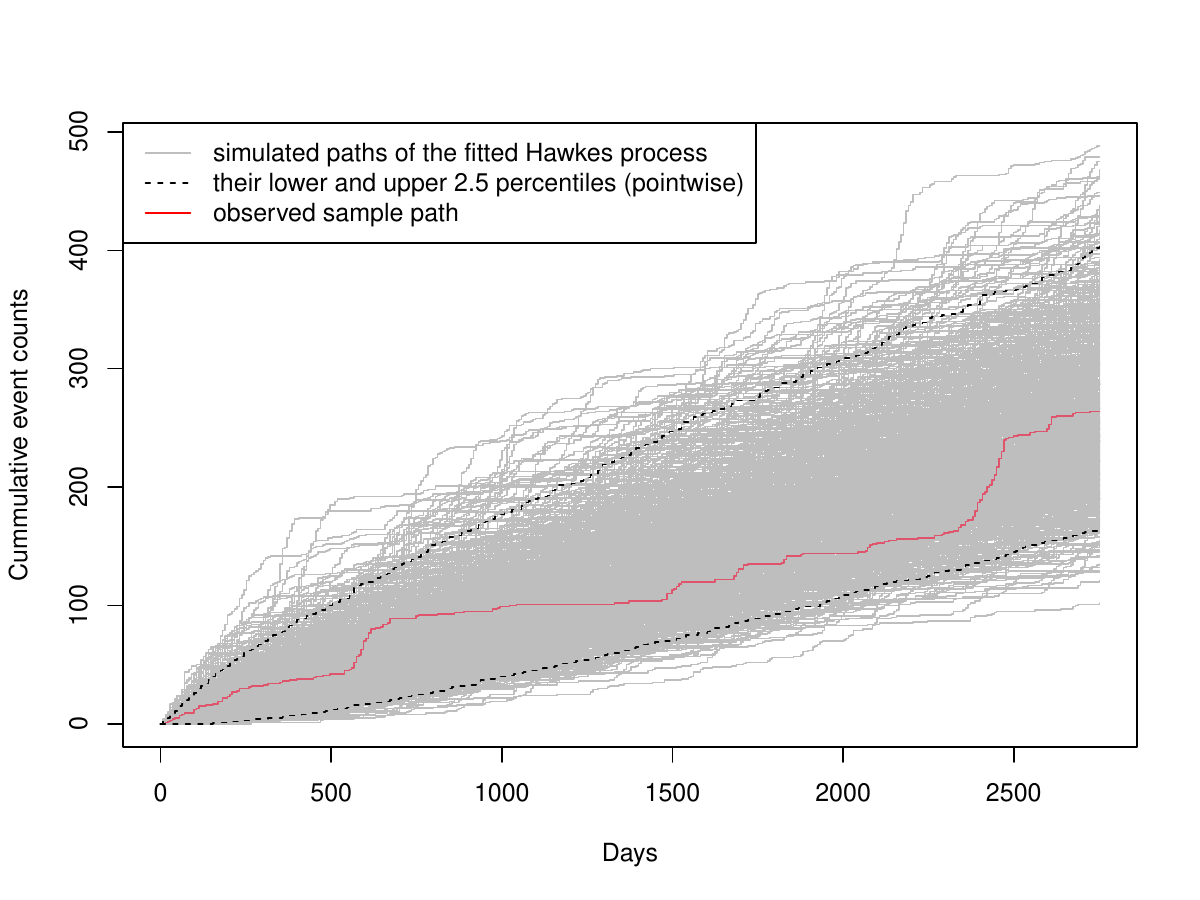}
  \includegraphics[width=0.49\textwidth]{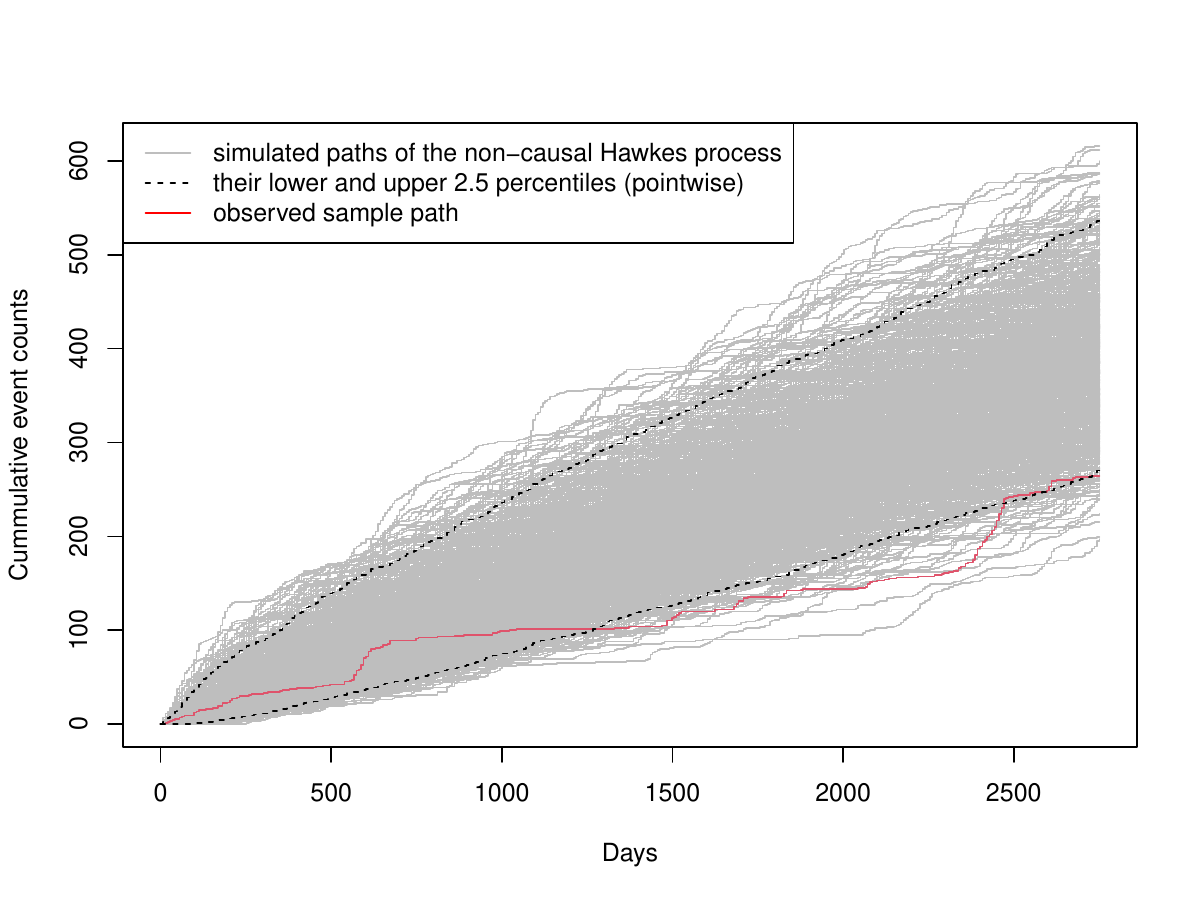} 
  \caption[Simulated paths of the fitted models]{Simulated sample
    paths of the fitted Hawkes process models together with the actual
    path. Left: the Hawkes process with an exponential kernel;
    right: the (non-causal) Hawkes process with a Gaussian kernel.}
\label{fig:simPaths}
\end{figure}

To assess whether the fitted Hawkes process model is adequate for the
observed data, we simulated 1000 times the discretely observed sample
path of the fitted model and compare the simulated paths with the
actual sample path implied by the data.  The left panel of
Figure~\ref{fig:simPaths} shows the 1000 simulated paths of the fitted
model together with the observed path. We note that the actual path
stays between the (pointwise) lower and upper 2.5 percentiles of the
simulated paths, suggesting the fitted model is adequate for the
weekly measles case count data. For comparison, we also simulated 1000
paths of the non-causal Hawkes process of \citeauthor{Cheysson2022}
and graph them together with the actual path in the right panel of
Figure~\ref{fig:simPaths}. The actual path wanders out of the 2.5
percentile lines for a substantial amount of time, suggesting the
non-causal Hawkes process is not adequate for the observed data. The
comparison suggests that the classical Hawkes process with an
exponential excitation kernel can offer adequate and better fit to the
data than the non-causal Hawkes process, despite having one parameter
less.

To entertain the possibility a kernel with a peak away from 0 might
fit the data better than an exponential kernel, we also fitted Hawkes
processes with Weibull and gamma density kernels to the data. In
addition to the scale parameter $\beta$, both kernels have an extra
parameter $\alpha\ (>0)$ that regulates the shape of the kernel. When
the shape parameter $\alpha$ is bigger than 1, both kernels will have
a peak away from zero; when the shape parameter is equal to 1, both
kernels reduce to the exponential kernel; and when the shape parameter
is less than 1, both kernels are decreasing on the positive real line
and approach infinity at zero. The fitted Hawkes processes with both
kernels have a shape parameter smaller than 1, with very marginal
improvement in the log-likelihood value. For example, with the Weibull
kernel
$\alpha(\cdot/\beta)^{\alpha-1}
\exp(-(\cdot/\beta)^\alpha)1\{\cdot>0\}$, the parameter estimates are
$\hat\nu=0.0204\ ([0.0112,0.0303])$,
$\hat\eta=0.788\ ([0.0644,0.903])$,
$\hat\alpha=0.742\ ([0.518,1.061]])$, and
$\hat\beta=9.43\ ([6.29,14.31])$, and the log-likelihood value at
$(\hat\nu,\hat\eta,\hat\alpha,\hat\beta)$ is $-379.83$
$([-380.10,-379.54])$, which is only a very minor improvement from the
log-likelihood value of $-380.19$ $([-380.47, -379.90])$ for the
fitted model with the exponential kernel. Therefore, there does not
appear to be statistical evidence to support a kernel with a peak away
from zero for the Hawkes process model. 

\section{Discussion}
\label{sec:discussion}

The main contribution of our work is the proposal of an unbiased
sequential Monte Carlo estimator of the intractable likelihood of the
Hawkes process relative to a discretely observed sample path and an
illustration of its usefulness in estimating the parameters of the
Hawkes process from such data when combined with the pseudo marginal
Metropolis-Hastings(PMMH) algorithm. In our numerical experiments, the
resulting estimator outforms two competitive estimators proposed in
the literature in terms of mean square error and behaves much like the
maximum likelihood estimator (MLE). Our method also gives the standard
errors of the estimators as a by-product.

While the pseudo marginal Metropolis-Hastings (PMMH) algorithm works
very well in our numerical experiments with farily arbitrary choices
for the number of particles and the step size of the Gaussian random
walk proposal, it is likely that computationally more efficient
procedures can be obtained by following the advices in the literature
on how to choose these tuning parameters optimally, such as
\cite{Pitt2012,Doucet2015,Gelman1997}.  Moreover, there have also been
more efficient variants of the PMMH algorithms such as the correlated
pseudomarginal method of \cite{Deligiannidis2018} and the unbiased
Markov Chain Monte Carlo method of \cite{Middleton2020}. Improvement
of the computational efficiency of our estimation procedure will is
interesting future work.

Justification of the use of the centre of the likelihood distribution
or more generally, the posterior distribution in a Bayesian framework,
for the model parameter to approximate the MLE relies on the
consistency and asymptotically normality expected of the MLE. While
there is strong empirical evidence to support these expected
properties of the MLE in the current context, a formal proof of such
properties is needed and shall be pursued elsewhere. 
\\[5pt]
\noindent \textbf{Acknowledgement} This work has benefited from
discussions with with Prof Pierre del Moral and Prof Judith Rousseau,
for which we are grateful. Part of the work was done while FC was
visiting the University of Oxford on a sabbatical trip. The
hospitality of Prof Samuel Cohen and Prof Judith Rousseau, the
Oxford-Man Institute of Quantitative Finance, the Mathematical
Institute and the Department of Statistics is gratefully
acknowledged. We also wish to thank Prof Xiaoli Meng for bringing
relevant references to our attention. This research includes
computations using the computational cluster Katana supported by
Research Technology Services at UNSW Sydney, as well as resources and
services at the National Computational Infrastructure and the Pawsey
Supercomputing Centre, both supported by the Australian Government’s
National Collaborative Research Infrastructure Strategy (NCRIS). This
research was supported by the Australian Government through the
Australian Research Council
(project number DP240101480).\\[12pt]


\appendix
\noindent{\textbf{\Large Appendices}}\nopagebreak
\section{Proof of the unbiasedness of the bootstrap particle
  approximation of the likelihood function}
\begin{proof}
  For ease of presentation, we use slightly more generic notation
  here. Let $y_{1:n}=(y_1,\dotsc,y_n)$ denote the time series of
  observations and $x_{1:n}=(x_1,\dotsc,x_n)$ the sequence of hidden
  states. Note that the $y_i$'s can have different dimensions, and so
  can the $x_i$'s. The bootstrap particle approximation to the likelihood
  $p(y_{1:n})$ presented in this work takes the form:
  \begin{align*}
    \hat p(y_{1:n})&=\prod_{i=1}^n \hat p(y_i|y_{1:i-1})\equiv \prod_{i=1}^n \frac1J \sum_{j=1}^J W_i^{(j)},
  \end{align*}
  where
  \begin{align*}
    W_i^{(j)}= p(y_i|y_{1:i-1}, x_{1:i, i}^{(j)})
    \bigg[\frac{P(\rmd x_i|x_{1:i-1,i}^{(j)}, y_{1:i-1})}
    {Q(\rmd x_i|x_{1:i-1, i}^{(j)},y_{1:i})} \bigg]_{x_i=x_{i,i}^{(j)}},
  \end{align*}
  and 
  \begin{align*}
    x_{1:i,i}^{(1:J)}\stackrel{\iid}{\sim}
    \frac{\frac 1 J \sum_{j=1}^J W_{i-1}^{(j)} \delta_{x_{1:i-1,i-1}^{(j)}}(\rmd x_{1:i-1})} {\hat p(y_{i-1}|y_{1:i-2})} Q(\rmd x_i|x_{1:i-1}, y_{1:i-1}).
  \end{align*}

  By our definition of the SMC approximation and the tower property of conditional expectations,
  \begin{align*}
    &\E{\hat p(y_i|y_{1:i-1})\big|x_{1:l,l}^{(1:J)}, l=1,\dotsc,i-1} \\
    {}={}&\E{\hat p(y_i|y_{1:i-1})\big|x_{1:i-1,i-1}^{(1:J)}}\\
    {}={}&\E{\E{\hat p(y_i|y_{1:i-1})\big|x_{1:i-1,i}^{(1:J)},x_{1:i-1,i-1}^{(1:J)}}\Big|x_{1:i-1,i-1}^{(1:J)}}\\
    {}={}&\E{\E{\frac 1 J \sum_{j=1}^J p(y_i|y_{1:i-1},x_{1:i,i}^{(j)}) 
       \Big[\frac{P(\rmd x_i|x_{1:i-1,i}^{(j)},y_{1:i-1})}{Q(\rmd x_i|x_{1:i-1,i}^{(j)},y_{1:i})}\Big]_{x_i=x_{i,i}^{(j)}}
       \Big|x_{1:i-1,i}^{(1:J)},x_{1:i-1,i-1}^{(1:J)}}\bigg|x_{1:i-1,i-1}^{(1:J)}}\\
    {}={}&\E{\frac 1 J \sum_{j=1}^J \E{p(y_i|y_{1:i-1},x_{1:i,i}^{(j)}) 
       \Big[\frac{P(\rmd x_i|x_{1:i-1,i}^{(j)},y_{1:i-1})}{Q(\rmd x_i|x_{1:i-1,i}^{(j)},y_{1:i})}\Big]_{x_i=x_{i,i}^{(j)}}
       \Big|x_{1:i-1,i}^{(1:J)},x_{1:i-1,i-1}^{(1:J)}}\bigg|x_{1:i-1,i-1}^{(1:J)}}\\
    {}={}&\E{\frac 1 J \sum_{j=1}^J \int p(y_i|y_{1:i-1},x_{1:i-1,i}^{(j)}, x_i) 
       \Big[\frac{P(\rmd x_i|x_{1:i-1,i}^{(j)},y_{1:i-1})}{Q(\rmd x_i|x_{1:i-1,i}^{(j)},y_{1:i})}\Big] Q(\rmd x_i|x_{1:i-1,i}^{(j)},y_{1:i})
       \bigg|x_{1:i-1,i-1}^{(1:J)}}\\
    {}={}&\E{\frac 1 J \sum_{j=1}^J \int p(y_i|y_{1:i-1},x_{1:i-1,i}^{(j)}, x_i) 
       P(\rmd x_i|x_{1:i-1,i}^{(j)},y_{1:i-1})
       \bigg|x_{1:i-1,i-1}^{(1:J)}}\\
    {}={}&\E{\frac 1 J \sum_{j=1}^J p(y_i|y_{1:i-1},x_{1:i-1,i}^{(j)})
       \big|x_{1:i-1,i-1}^{(1:J)}}. 
  \end{align*}
  Conditional on $x_{1:i-1,i-1}^{(1:J)}$, the $x_{1:i-1,i}^{(j)},\ j=1,\dots,J$ are
  \iid\ multinomial and each of them takes the value $x_{1:i-1,i-1}^{(j)}$ with
  probability $\frac 1 J W_{i-1}^{(j)}/\hat p(y_{i-1}|y_{1:i-2})$, $j=1,\dotsc,J$. Therefore,
  \begin{align*}
    &\E{\hat p(y_i|y_{1:i-1})\big|x_{1:l,l}^{(1:J)}, l=1,\dotsc,i-1}\\
    {}={}&\E{p(y_i|y_{1:i-1},x_{1:i-1,i}^{(1)})
       \big|x_{1:i-1,i-1}^{(1:J)}}\\
    {}={}&\frac{\frac 1 J \sum_{j=1}^J p(y_i|y_{1:i-1}, x_{1:i-1,i-1}^{(j)}) W_{i-1}^{(j)} }{\hat p(y_{i-1}|y_{1:i-2})}\\
    {}={}&\frac{\frac 1 J \sum_{j=1}^J p(y_i|y_{1:i-1}, x_{1:i-1,i-1}^{(j)}) p(y_{i-1}|y_{1:i-2}, x_{1:i-1,i-1}^{(j)})\Big[\frac{P(\rmd x_{i-1}|x_{1:i-2,i-1}^{(j)},y_{1:i-2})}{Q(\rmd x_{i-1}|x_{1:i-2,i-1}^{(j)},y_{1:i-1})}\Big]_{x_{i-1}=x_{i-1,i-1}^{(j)}} }{\hat p(y_{i-1}|y_{1:i-2})}\\
    {}={}&\frac{\frac 1 J \sum_{j=1}^J p(y_{i-1:i}|y_{1:i-2}, x_{1:i-1,i-1}^{(j)})\Big[\frac{P(\rmd x_{i-1}|x_{1:i-2,i-1}^{(j)},y_{1:i-2})}{Q(\rmd x_{i-1}|x_{1:i-2,i-1}^{(j)},y_{1:i-1})}\Big]_{x_{i-1}=x_{i-1,i-1}^{(j)}} }{\hat p(y_{i-1}|y_{1:i-2})}. 
  \end{align*}
  From this, and by double expectation, we have
  \begin{align*}
    &\E{\hat p(y_{1:n})}\\
    {}={}&\E{\E{\prod_{i=1}^n\hat p(y_i|y_{1:i-1})\Big|x_{1:l,l}^{(1:J)}, l=1,\dotsc,n-1 } }\\
    {}={}&\E{\prod_{i=1}^{n-1}\hat p(y_i|y_{1:i-1}) \E{\hat p(y_n|y_{1:n-1})\Big|x_{1:n-1,n-1}^{(1:J)}} }\\
    {}={}&\E{\prod_{i=1}^{n-2}\hat p(y_i|y_{1:i-1}) \frac 1 J \sum_{j=1}^J p(y_{n-1:n}|y_{1:n-2}, x_{1:n-1,n-1}^{(j)})\Big[\frac{P(\rmd x_{n-1}|x_{1:n-2,n-1}^{(j)},y_{1:n-2})}{Q(\rmd x_{n-1}|x_{1:n-2,n-1}^{(j)},y_{1:n-1})}\Big]_{x_{n-1}=x_{n-1,n-1}^{(j)}}}.
  \end{align*}
  Now, by conditioning on $x_{1:l,l}^{(1:J)}, l=1,\dotsc,n-2$, we have
  \begin{align*}
    &\E{\hat p(y_{1:n})}\\
    {}={}& \E{\prod_{i=1}^{n-3}\hat p(y_i|y_{1:i-1}) \frac 1 J \sum_{j=1}^J p(y_{n-2:n}|y_{1:n-3}, x_{1:n-2,n-2}^{(j)})\Big[\frac{P(\rmd x_{n-2}|x_{1:n-3,n-2}^{(j)},y_{1:n-3})}{Q(\rmd x_{n-2}|x_{1:n-3,n-2}^{(j)},y_{1:n-2})}\Big]_{x_{n-2}=x_{n-2,n-2}^{(j)}}}.
  \end{align*}
  Repeating this process, we have
  \begin{align*}
    &\E{\hat p(y_{1:n})}\\
    {}={}&\E{\frac 1 J \sum_{j=1}^J p(y_{1:n}|x_{1,1}^{(j)})\Big[\frac{P(\rmd x_1)}{Q(\rmd x_1|y_1)}\Big]_{x_1=x_{1,1}^{(j)}}}\\
    {}={}&\E{p(y_{1:n}|x_{1,1}^{(1)})\Big[\frac{P(\rmd x_1)}{Q(\rmd x_1|y_1)}\Big]_{x_1=x_{1,1}^{(1)}}}\\
    {}={}& \int p(y_{1:n}|x_1) \Big[\frac{P(\rmd x_1)}{Q(\rmd x_1|y_1)}\Big] Q(\rmd x_1|y_1)\\
    {}={}& \int p(y_{1:n}|x_1) P(\rmd x_1)\\
    {}={}& p(y_{1:n}). \qedhere
  \end{align*}
\end{proof}

\section{Proof of the exactness of the MCMC approximation to the
  likelihood distribution}
\begin{proof}
  Write the SMC likelihood approximation $\hat L(\theta,u)$ to remind
  us of its dependence on the Monte Carlo randomness $u$, which can be
  taken as a uniform random variable on the interval $(0,1)$ \citep[see
  e.g.][Lemma~4.22]{Kallenberg2021}. The Markov chain
  $\theta^{(1)}, \theta^{(2)}, \dotsc$ we constructed using the PMMH
  algorithm to approximate the likelihood distribution
  $L(\theta)\rmd \theta/\int L(\theta)\rmd \theta$ can be viewed as
  the second to last dimension of the Markov chain
  $(u^{(i)},\theta^{(i)})$, $i=1,2,\dotsc$ with transition kernel 
  \begin{align*}
    K((\rmd u',\rmd \theta')|(u,\theta))=\rmd u'Q(\rmd \theta'|\theta)
    \brackets{\frac{\hat L(\theta',u')\rmd \theta' Q(\rmd
    \theta|\theta')}{\hat L(\theta,u)\rmd \theta Q(\rmd
    \theta'|\theta)}\wedge 1}+R(u,\theta) \delta_{(u,\theta)}(\rmd u', \rmd \theta'),
  \end{align*}
  where $x\wedge y=\min\set{x,y}$ and
  $R(u,\theta)=1-\iint \brackets{\frac{\hat L(\theta',u')\rmd \theta'
      Q(\rmd \theta|\theta')}{\hat L(\theta,u)\rmd \theta Q(\rmd
      \theta'|\theta)}\wedge 1} \rmd u'Q(\rmd \theta'|\theta) $. 
  For $(u,\theta)\neq (u',\theta')$, we note
  \begin{align*}
    &\rmd u\, \hat L(\theta,u)\rmd\theta\, K((\rmd u',\rmd
      \theta')|(u,\theta))\\
    {}={}& \rmd u\, \hat L(\theta,u)\rmd\theta \rmd u'Q(\rmd \theta'|\theta)
           \brackets{\frac{\hat L(\theta',u')\rmd \theta' Q(\rmd
           \theta|\theta')}{\hat L(\theta,u)\rmd \theta Q(\rmd
           \theta'|\theta)}\wedge 1}\\
    {}={}& \rmd u\, \rmd u' \hat L(\theta',u')\rmd \theta' Q(\rmd
           \theta|\theta') \ \wedge\ \rmd u\, \hat
           L(\theta,u)\rmd\theta \rmd u'Q(\rmd \theta'|\theta)\\
    {}={}& \rmd u'\, \hat L(\theta',u')\rmd\theta'\, K((\rmd u,\rmd
      \theta)|(u',\theta')),
  \end{align*}
  and therefore
  $\rmd u \hat L(\theta,u)\rmd \theta / \iint \hat L(\theta,u)\rmd
  \theta\rmd u $ is the invariant distribution of the extended chain
  $(u^{(i)},\theta^{(i)}), i=1,2,\dotsc$. The $\theta$-marginal of
  this invariant distribution is
  \begin{align*}
  \brackets{\int \hat L(\theta,u)\rmd u}\rmd \theta \Big/ \iint \hat
  L(\theta,u)\rmd \theta\rmd u= L(\theta)\rmd \theta \Big/\int L(\theta)\rmd \theta
  \end{align*}
  by Proposition~\ref{prop:unbiasedness}. This shows that, as the
  $\theta$-marginal of the extended chain,
  $\theta^{(i)}, i=1,2,\dotsc,$ has the likelihood distribution as its
  invariant distribution. 
\end{proof}
\bibliographystyle{apalike}
\bibliography{ntvhpest}

\begin{thebibliography}{}

\bibitem[Andrieu et~al., 2010]{Andrieu2010}
Andrieu, C., Doucet, A., and Holenstein, R. (2010).
\newblock Particle markov chain monte carlo methods.
\newblock {\em Journal of the Royal Statistical Society: Series B (Statistical
  Methodology)}, 72(3):269--342.

\bibitem[Andrieu and Roberts, 2009]{Andrieu2009}
Andrieu, C. and Roberts, G.~O. (2009).
\newblock {The pseudo-marginal approach for efficient Monte Carlo
  computations}.
\newblock {\em The Annals of Statistics}, 37(2):697 -- 725.

\bibitem[Bezanson et~al., 2017]{Bezanson2017julia}
Bezanson, J., Edelman, A., Karpinski, S., and Shah, V.~B. (2017).
\newblock Julia: A fresh approach to numerical computing.
\newblock {\em SIAM review}, 59(1):65--98.

\bibitem[Chen, 2022]{Chen2022a}
Chen, F. (2022).
\newblock {\em IHSEP: Inhomogeneous Self-Exciting Process}.
\newblock R package version 0.3.1. URL:
  \url{https://cran.r-project.org/package=IHSEP}.

\bibitem[Cheysson, 2021]{Cheysson2021Rpackage}
Cheysson, F. (2021).
\newblock {\em hawkesbow: Estimation of Hawkes Processes from Binned
  Observations}.
\newblock R package version 1.0.2.

\bibitem[Cheysson and Lang, 2022]{Cheysson2022}
Cheysson, F. and Lang, G. (2022).
\newblock Spectral estimation of {H}awkes processes from count data.
\newblock {\em The Annals of Statistics}, 50(3):1722 -- 1746.

\bibitem[Chornoboy et~al., 1988]{Chornoboy1988}
Chornoboy, E., Schramm, L., and Karr, A. (1988).
\newblock {Maximum likelihood identification of neural point process systems}.
\newblock {\em Biological Cybernetics}, 59(4):265--275.

\bibitem[Daley and Vere-Jones, 2003]{Daley2003}
Daley, D.~J. and Vere-Jones, D. (2003).
\newblock {\em An Introduction to the Theory of Point Processes Volume I:
  Elementary Theory and Methods}.
\newblock Springer-Verlag, New York, 2nd edition.

\bibitem[Deligiannidis and Doucet, 2018]{Deligiannidis2018}
Deligiannidis, G. and Doucet, A. (2018).
\newblock The correlated pseudomarginal method.
\newblock {\em Journal of the Royal Statistical Society. Series B (Statistical
  Methodology)}, 80(5):pp. 839--870.

\bibitem[Doucet et~al., 2015]{Doucet2015}
Doucet, A., Pitt, M.~K., Deligiannidis, G., and Kohn, R. (2015).
\newblock Efficient implementation of {M}arkov chain {M}onte {C}arlo when using
  an unbiased likelihood estimator.
\newblock {\em Biometrika}, 102(2):295--313.

\bibitem[Gelman et~al., 1997]{Gelman1997}
Gelman, A., Gilks, W.~R., and Roberts, G.~O. (1997).
\newblock {Weak convergence and optimal scaling of random walk Metropolis
  algorithms}.
\newblock {\em The Annals of Applied Probability}, 7(1):110 -- 120.

\bibitem[Gordon et~al., 1993]{GordonEtAl1993}
Gordon, N., Salmond, D., and Smith, A. (1993).
\newblock Novel approach to nonlinear/non-{G}aussian {B}ayesian state
  estimation.
\newblock {\em IEE Proceedings F - Radar and Signal Processing},
  140:107--113(6).

\bibitem[Hastings, 1970]{Hastings1970}
Hastings, W.~K. (1970).
\newblock {M}onte {C}arlo sampling methods using {M}arkov chains and their
  applications.
\newblock {\em Biometrika}, 57(1):97--109.

\bibitem[Hawkes, 1971]{Hawkes1971}
Hawkes, A.~G. (1971).
\newblock Spectra of some self-exciting and mutually exciting point processes.
\newblock {\em Biometrika}, 58(1):83--90.

\bibitem[Kallenberg, 2021]{Kallenberg2021}
Kallenberg, O. (2021).
\newblock {\em Foundations of Modern Probability}.
\newblock Springer Nature, Switzerland, 3rd edition.

\bibitem[Kitagawa, 1996]{Kitagawa1996}
Kitagawa, G. (1996).
\newblock Monte carlo filter and smoother for non-gaussian nonlinear state
  space models.
\newblock {\em Journal of Computational and Graphical Statistics}, 5(1):1--25.

\bibitem[Metropolis et~al., 1953]{Metropolis1953}
Metropolis, N., Rosenbluth, A.~W., Rosenbluth, M.~N., Teller, A.~H., and
  Teller, E. (1953).
\newblock Equation of state calculations by fast computing machines.
\newblock {\em The Journal of Chemical Physics}, 21(6):1087--1092.

\bibitem[Middleton et~al., 2020]{Middleton2020}
Middleton, L., Deligiannidis, G., Doucet, A., and Jacob, P.~E. (2020).
\newblock {Unbiased Markov chain Monte Carlo for intractable target
  distributions}.
\newblock {\em Electronic Journal of Statistics}, 14(2):2842 -- 2891.

\bibitem[Nishiura et~al., 2017]{Nishiura2017}
Nishiura, H., Mizumoto, K., and Asai, Y. (2017).
\newblock Assessing the transmission dynamics of measles in japan, 2016.
\newblock {\em Epidemics}, 20:67--72.

\bibitem[Ogata, 1978]{Ogata1978}
Ogata, Y. (1978).
\newblock The asymptotic behaviour of maximum likelihood estimators for
  stationary point processes.
\newblock {\em Annals of the Institute of Statistical Mathematics},
  30:243--261.

\bibitem[Ozaki, 1979]{Ozaki1979}
Ozaki, T. (1979).
\newblock Maximum likelihood estimation of {H}awkes' self-exciting point
  processes.
\newblock {\em Annals of the Institute of Statistical Mathematics},
  31(1):145--155.

\bibitem[Pitt et~al., 2012]{Pitt2012}
Pitt, M.~K., dos Santos~Silva, R., Giordani, P., and Kohn, R. (2012).
\newblock On some properties of markov chain monte carlo simulation methods
  based on the particle filter.
\newblock {\em Journal of Econometrics}, 171(2):134--151.
\newblock Bayesian Models, Methods and Applications.

\bibitem[Pitt and Shephard, 1999]{PittShephard1999}
Pitt, M.~K. and Shephard, N. (1999).
\newblock Filtering via simulation: Auxiliary particle filters.
\newblock {\em Journal of the American Statistical Association},
  94(446):590--599.

\bibitem[{R Core Team}, 2022]{RCoreTeam2022}
{R Core Team} (2022).
\newblock {\em R: A Language and Environment for Statistical Computing}.
\newblock R Foundation for Statistical Computing, Vienna, Austria.

\bibitem[Rizoiu et~al., 2022]{Rizoiu2022}
Rizoiu, M.-A., Soen, A., Li, S., Calderon, P., Dong, L.~J., Menon, A.~K., and
  Xie, L. (2022).
\newblock Interval-censored hawkes processes.
\newblock {\em Journal of Machine Learning Research}, 23(338):1--84.

\bibitem[Shcherbinin, 1987]{Shcherbinin1987}
Shcherbinin, A.~F. (1987).
\newblock The normalized likelihood method.
\newblock {\em Measurement Techniques}, 30(12):1129--1134.

\bibitem[Shlomovich et~al., 2022a]{Shlomovich2022}
Shlomovich, L., Cohen, E. A.~K., and Adams, N. (2022a).
\newblock A parameter estimation method for multivariate binned {H}awkes
  processes.
\newblock {\em Statistics and Computing}, 32(6):98.

\bibitem[Shlomovich et~al., 2022b]{Shlomovich2022jcgs}
Shlomovich, L., Cohen, E. A.~K., Adams, N., and Patel, L. (2022b).
\newblock Parameter estimation of binned {H}awkes processes.
\newblock {\em Journal of Computational and Graphical Statistics},
  31(4):990--1000.

\bibitem[{van der Vaart}, 2007]{vandervaart07}
{van der Vaart}, A. (2007).
\newblock {\em Asymptotic Statistics}.
\newblock Cambridge university press, New York.

\bibitem[Xie and Singh, 2013]{Xie2013}
Xie, M.-g. and Singh, K. (2013).
\newblock Confidence distribution, the frequentist distribution estimator of a
  parameter: A review.
\newblock {\em International Statistical Review}, 81(1):3--39.

\end{thebibliography}
\end{document}